\documentclass[a4paper,twocolumn]{article}

\usepackage{graphicx}                 
\usepackage{epsfig, subfigure}
\usepackage{amsmath, amsthm, amssymb} 

\usepackage[latin1]{inputenc}         
\usepackage{enumerate}

\begin{document}

\title{Conservation laws, financial entropy  and the Eurozone crisis}
\author{Paul Cockshott and David Zachariah\footnote{E-mail:
    \texttt{William.Cockshott@glasgow.ac.uk} and \texttt{dave.zachariah@gmail.com}}}

\maketitle
\begin{abstract}
The report attempts of apply econophysics concepts
to the Eurozone crisis. It starts by examining the idea of
conservation laws as applied to market economies.
It formulates a measure of financial entropy and gives
numerical simulations indicating that this tends to rise.
We discuss an analogue for free energy released
during this process.
The concepts of real and symbolic appropriation
are introduced as a means to analyse debt and
taxation.

We then examine the conflict between the conservation
laws that apply to commodity exchange with the
exponential growth implied by capital accumulation
and how these have necessitated a sequence of
evolutionary forms for money,
and go on to present a simple stochastic model for the
formation of rates of interest and a model for the time
evolution of the rate of profit.

Finally we apply the conservation law model to examining
the Euro Crisis and the European Stability pact, arguing that
if the laws we hypothesise actually hold, then the
goals of the stability pact are unobtainable.
\end{abstract}
\tableofcontents{}



\section{Conservation laws in market exchange}

In the physical sciences conservation laws provide a basic framework
within which theories are cast. They are so fundamental that it is
hard to conceive of any mathematically expressed system of mechanics that does not rest on such laws.

Do similar laws exist in economic theory?

In a sense yes, where they are termed `accounting
identities' and regarded as something pretty trivial. We want to argue
that this view of them as something trivial is misplaced, and that
they can actually tell us a lot more about the nature of social relations
and their degree of constrainedness than is generally realised.

Marx, to an extent greater than is sometimes recognised, wanted to
establish a theory of the capitalist economy informed by the laws
of physics. \cite[ch.~2]{Banaji2010_theory} This comes across in several
ways: his avowed aim to write a book on the `laws of motion' of
capitalism; his distinction between the concept of labour and labour
power; his presentation of value as the crystalisation of human
energy; and his analysis of commodity exchange as an equivalence
relation.

Marx said that in \emph{Capital} he was investigating the `laws of motion'
of capitalism. This might be understood as only a metaphor derived
from physics, but we think that it is worth taking it seriously. If
you think of the time he was writing -- the 1860s -- one of the key recent
discoveries of physics was the idea of the conservation of energy.
The conservation of energy had been formalised by Helmholtz and Grove
in the late 1840s. This held that although energy might appear in
various forms: heat, motion, gravitational potential, it was conserved
in its exchange between these forms.\footnote{``What Lucretius says is self-evident; `nil posse creari de nihilo,' out of nothing, nothing can be
created. Creation of value is transformation of labour-power into labour. Labourpower itself is energy
transferred to a human organism by means of nourishing matter.'' \cite[ch.~9]{Marx_vol1}}

Marx's initial argument in \emph{Capital}, before he derives labour power
as the source of surplus value is similar. Value is neither created
nor destroyed in the exchange process, but can only change its form.
His argument asserts in effect a law of the conservation of value
in exchange.

Think of the distinction between `labour' and `labour power'. This is
so closely parallels  Watt's distinction between `work' and `power',
that it is surprising that the similarity is rarely remarked on.\footnote{This need not have been via a direct study of
Watt by Marx. Watt's distinction between work done and
power: the ability to perform work, measured in standard
horse-powers had become a commonplace of industrial
society.}
His analysis of commodity exchange is also structured like an analysis
of a conservation law \cite{Marx_vol1}. He introduces as an example
\begin{center}
20 yards of linen = 1 coat or 20 yards of linen are worth 1 coat,
\end{center}
in this notation he says that the coat plays the role of the equivalent
and that it implies also the converse relation
\begin{center}
1 coat = 20 yards of linen or 1 coat is worth 20 yards of linen.
\end{center}
He then presents what he calls the expanded form of the relation

20 yards of linen = 1 coat

20 yards of linen = 10 lbs of tea, etc.

And goes on from this to state that 1 coat will be equal to 10 lbs
tea. What he is doing here is setting out what in modern mathematical
terminology is an equivalence relation. For some relational operator
$\doteqdot$ we say that $\doteqdot$ is an equivalence relation if
the relations is commutative, transitive and reflexive, that is
\begin{itemize}
\item if $a\doteqdot a$

\item and if $a\doteqdot b$ implies $b\doteqdot a$

\item and if $a\doteqdot b$ and $b\doteqdot c$ implies $a\doteqdot c$
\end{itemize}
then the relation $\doteqdot$ is an equivalence relation. In his case
$\doteqdot$ would stand for exchange of commodities. Now equivalence
relations are interesting because systems goverened by conservation
laws display them. Thus in a many-body gravitational problem with
a predefined collection of particle masses, the set of possible configurations
of particle position and velocities is partitioned into equivalence
sets with respect to energy. Within each set all configurations share
the same total energy and the conservation of energy prevents transitions
of configurations between these sets.

This is in essence what we understand by a conservation law.

We infer the existence of energy as a conserved quantity by the fact
that, in closed systems, we never observe transitions between configurations
with different total energies.

Marx's demonstration that commodity exchange is an equivalence relation
is then used to infer that there is a conserved quantity `value', the
sum of which is unchanged under the operation of exchanges. Commodity
exchange is governed by a conservation law. To borrow
much later terminology, exchange is shown to
be a zero-sum game.

This may seem a trivial observation, but it leads to an important
deduction: that in a conservative system, any surplus of value --
profit -- must arise from outside of the system and thus that profit
must originate from production rather than exchange. Marx argued that
the conserved quantity in commodity exchanges is human energy expended
as labour and that this energy provides the external input that allows
a surplus. At the economy's two ends, production and consumption,
the process is nonconservative but in between them lies  market
exchange: a conservative system.

\subsection{Finance and conservation laws}

Although economists are sometimes dismissive of conservation laws, disparagingly called `accounting identities',
these identities can still throw useful light on the financial crisis
that has been unfolding these last few years. But to do this we need
to uncover the specific laws of motion, that is to say both the conservation
laws and the particle dynamics that govern the financial system.

That we have to think of the financial system using tools derived
from statistical mechanics should be obvious by now. It is over a quarter of
a century since it was shown that the regulation of prices by labour
content arises directly from statistical mechanical considerations \cite{Farjoun&Machover1983}.

If we look at the more recent work on the statistical
mechanics of money \cite{Dragulescu&Yakovenko2000,Dragulescu&Yakovenko2002_survey}, we can see a similar
structure of argumentation -- pushed somewhat by the use of more
sophisticated physics. This research argues that if we
treat trades between commodity owners as a random process that conserves
money, then you can use statistical mechanics to make deductions about
the distribution of money. Money as a conserved quantity randomly
transfered in exchanges between agents, models the transfer of energy
between gas molecules and it follows that the maximum entropy distribution
of money will be a Gibbs-Boltzmann one. The statistical
mechanics of money shows that this distribution fits the observed
distribution of money for most of the population, but that a small minority of very rich people fall off
this distribution -- their wealth follows a power-law. The Gibbs
distribution falls of sharply at high levels of wealth, whereas the
the tail of the power law distribution extends much
further.

The probability of anyone being as rich as Bill Gates or Warren Buffet
as a result of simple commodity trading is vanishingly small, so
\cite{Dragulescu&Yakovenko2001,Silva&Yakovenko2005} conclude that there has to be some mechanism outside of
equivalent exchange that gives rise to their extraordinary wealth: the
effect of compound interest which is a non-random process.

Research in agent-based modeling has shown that if you
partition the population into buyers and sellers of labour power and
run agent-based simulations, you a reproduce the combination of Gibbs
and power-law distributions of wealth \cite{Wright2005_socialarc}
observed in real capitalist economies
\cite{Dragulescu&Yakovenko2001,Silva&Yakovenko2005}. Further, when
producers can specialise in and switch between different
commodity-types, Marx's law of labour value appear as an emergent
phenomenon from local and distributed market exchanges \cite{Wright2008_emergence}.

\subsection{The phase space of finance}
A basic tool of conceptual analysis in statistical mechanics is the
concept of phase space. If you consider a collection of particles
(for example stars) in a closed volume, each particle can be described
by 6 numbers which specify its position and momentum in terms of a
three-dimensional Cartesian coordinate system: 3 numbers to specify the
position and 3 numbers to specify its momentum. We say that each particle
has 6 \emph{degrees of freedom}.

So if we have a million particles, for instance if one is considering
the dynamics of a galaxy, the system has 6 million degrees of freedom
and we can consider these to be a coordinate system such that every
possible configuration of molecular positions and moments constitutes
a point in this 6 million dimensional space. We call such a space
a \emph{phase space}. The laws of motion then specify a trajectory
of the whole system through this phase space.

 The overall system is
governed by conservation laws. The mass is conserved, and relative
to the center of gravity of the system as a whole the sum of the momenta
in each of the directions of our Cartesian system must sum to zero.

Why is this relevant to finance?

Well we are again dealing with a system with very large numbers of
agents and we have analogues of position and momentum. The total debt/credit
position of an agent is analogous to its mass, and the rate of change
of its debt/credit position is analogous to its momentum. Thus if
two billion people in the world are enmeshed in debt/credit relations
then the whole system can be thought of as a phase space of 2 billion
dimensions.

\begin{figure}[h]
  \begin{center}
    \includegraphics[width=0.9\columnwidth]{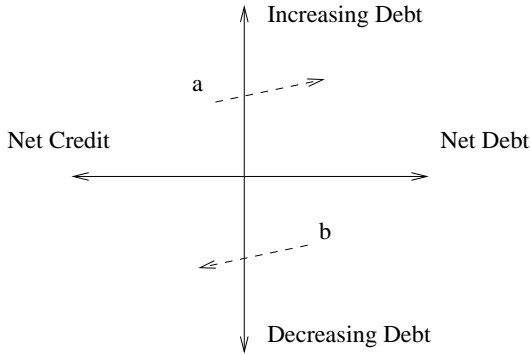}
  \end{center}
  \caption{A phase diagram for financial space relating debt to the
    rate of change of debt. Agent $a$ is currently a net creditor, but
    it is borrowing and as such moving to the right towards becoming
    a net debtor. Agent $b$ has debts which it is paying off and
    moving to the left to become a net creditor.}
\label{fig:phaseplane}
\end{figure}

It is impossible to visualise a space with high numbers of dimensions,
so there are graphical techniques that people use to reason about
them. One trick is to project the high dimensional phase space down
onto only two dimensions: for example, position and momentum in
the direction of one of the axes of the coordinate system. This allows one to create a phase diagram. Applied to the financial system, a possible phase diagram
is  shown in Figure \ref{fig:phaseplane}.

We could show every agent (firm, individual, state) as a point on
this plane. Or in a more abstract way we could say that there is a
density function $P_{x,y}$ which gives the probability of finding
an agent in a given area in the phase plane. Using this probablility
density function (PDF) we can formulate two conservation laws analogous to the conservation
of mass and momentum.
\begin{enumerate}[i.]
\item The total `mass' on the left of the origin equals the mass on the
right of the origin. Since the mass to the left is credit, and that
to the right is debt, this is another way of saying that the total
debt and total credit must balance.
\begin{equation}
0=\intop_{-\infty}^{\infty}xP_{x}dx\label{eq:conservationofmass}
\end{equation}

\item The total `mass' above the origin must equal the mass below the origin,
that is to say that any growth in debt must be compensated for by
a growth in credit. This is an `equal and opposite reaction' effect.
\begin{equation}
0=\int_{-\infty}^{\infty}yP_{y}dy\label{eq:conservationofmomentum}
\end{equation}

\end{enumerate}
These very basic points establish that there is no net value or net
wealth embodied in the financial system and that there is no flow
of value into or out of the financial system. It shows the fallacy
of the conception, popular among some political economists that capital
has `moved into finance' because of the low rate of profit pertaining
in industry.

This is a fundamental misconception, capital is value,
and it can not flow into the financial system, since the sum of value
here is always zero. A moment's thought about the materiality of value
confirms this. Real-economic value -- what the classical economists
understood to be the labour content of physical products
\cite{Ricardo1817} -- can not be converted into financial instruments which are just information structures.

A similar dimensional argument shows the fallacy of the idea of the `money supply' used in
orthodox economics. The notion of a supply originates with physical
flows like the supply of water to a town, or the supply of cars provided
by all the car factories in the world. There are flows in the financial
system, e.g. flows of agents towards greater debt, but these are exactly
balanced by flows towards greater credit positions, so the net flow
is zero. If the term money supply is taken instead to mean a stock
of money, and this is taken to include bank deposits, one then has
the problem that both negative and positive money exist: liabilities and
assets of the banking system respectively. Taken as a whole the
positive and negative stocks cancel.

We believe it is better to try and understand the system in terms
of its laws of motion. These laws are both the conservation laws \eqref{eq:conservationofmass}
and \eqref{eq:conservationofmomentum}, and the forces acting on the
individual particles (capitalist firms, states, etc.). We will argue that the
degree of `disorder' or entropy of a financial system tends to
increase over time. We would expect this of any chaotic system
governed by conservation laws, but an examination of the force fields
acting on the particles will show why this takes place.
\subsection{Tendancy of financial entropy to rise}

With the density function $P_{x,y}$ we can compute the entropy of a system, denoted $H$,  using the standard Boltzman formula,
\begin{equation}
H=-\int P_{x,y} \ln P_{x,y} \: dx dy,
\end{equation}
and we assert that for a financial system $H$ will rise over time.

Why does the entropy tend to increase?

First a common-sense explanation. Consider a collection of firms or
enterprises, distributed on the phase plane diagram as in
Figure~\ref{fig:phaseplane}. Each firm can be considered a
particle subject to forces which determine its rate of change of
debt. There are three cases to consider:
\begin{enumerate}[(A)]
\item A firm whose debt is rising because its profits are too low
even to meet its interest payments to the banks. Such a firm has to
borrow more from the bank to stay in operation. This firm is an
\emph{involuntary borrower}.
\item A firm may be \emph{voluntarily borrowing} because it has a high rate of
profit, well in excess of the interest rate, and thus can increase
its profit of enterprise by taking on bank loans to invest and expand its business.
\item A firm may reckon that the rate of profit it can earn is lower than
the rate of interest, but its current debt level may be quite low
so that it is in a position to pay off its debts to the bank with
some of its retained profit. A firm may even have no net borrowing
and find that it is more profitable to earn interest on its cash than
it is to invest it productively. Such a firm is a \emph{voluntary lender}.
\end{enumerate}
The total amount of borrowing must equal to the total amount
of lending, thus if the quantity of voluntary borrowing by firms in
group (B) falls below the quantity of lending by firms in group (C), the
lack of demand from investments will automatically create sufficient
firms in (A) trying to stay in operation through involuntary
borrowing, ensuring that equation \eqref{eq:conservationofmomentum} is
met. The net effect of this is to polarise capital in the phase plane
as borrowers and lenders, giving rise to a tadpole shape with a head
of productive firms and a tail of rentier firms, who earn their
revenue from financial rather than productive assets. Therefore, as
the dispersion or spread of the probability density function
increases, the entropy $H$ rises, representing more disorder.

\begin{figure}[h]
\begin{tabular}{ccc}
  $T$ &
PDF in the phase plane $P_{x,y}$ &
  $H$\tabularnewline
2 &
\includegraphics[width=0.7\columnwidth]{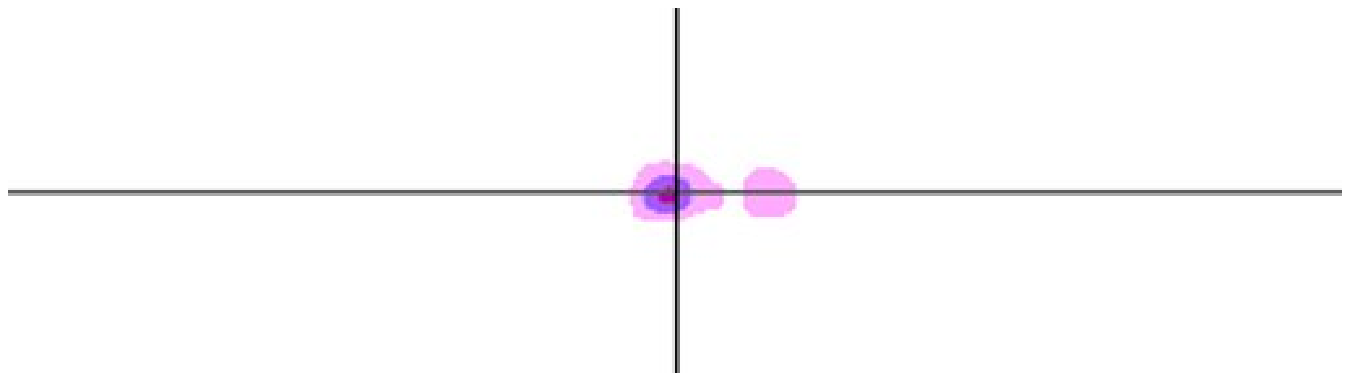} &
3.29\tabularnewline
5 &
\includegraphics[width=0.7\columnwidth]{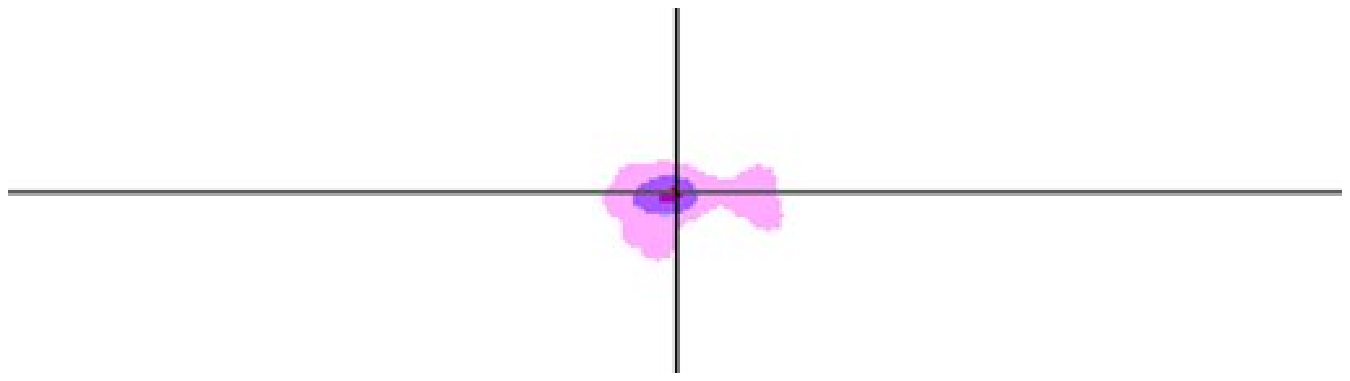} &
3.72\tabularnewline
20 &
\includegraphics[width=0.7\columnwidth]{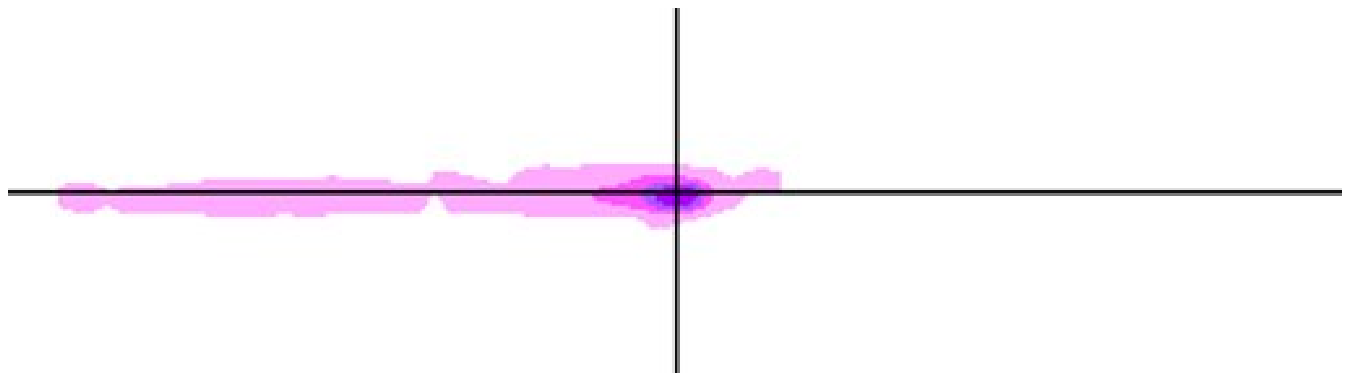} &
4.74\tabularnewline
\end{tabular}

\caption{The formation of a rentier tail due to the polarisation of capitals
in the phase plane. $T$ is timestep, $H$ is entropy. The images show a
plot of the probability density function of capitals in the phase
plane. The horizontal axis shows the net debt, normalised by the
total capital stock, i.e., the `debt ratio'. The vertical axis
shows the rate of change of the debt, also normalised by total capital
stock. As time passes the entropy of the system increases.
Note that the colour is nonlinearly related to the actual value of
$P_{x,y}$ in order to display the areas of very low but non-zero probability
in the tail.}
\label{fig:The-formation-of-rentier-tail}
\end{figure}

In previous work \cite{Cockshott&Cottrell2008_probabilistic} we
presented a numerical simulation model based on the methodology
developed in \cite{Wright2005_socialarc} that models a simple capitalist economy with a large number
of capitalists and workers. The state of indebtedness/credit of each
of the capitals is tracked as the economy evolves. The capitalists interact
via a very simple financial system represented by a single bank which
maintains debit and credit accounts for agents based on a certain
amount of base money which acts as its reserve. The evolution can be
visualised in a phase plane diagram. For clarity we normalise the
net debt of each firm by its total capital stock, which is known as the `debt ratio'.\footnote{Note that a firm with a net credit position has a
  negative debt ratio, $D/K$. The capital stock $K$ was calculated from the current value of the commodity stocks and equipment of the enterprise.} Similarly the rate of change of debt is
normalised by its total capital stock. Initially the capitalists are
clustered around the origin of the phase plane diagram, but as time
passes the distribution becomes elongated with head of relatively
indebted capitalists and long `rentier tail' of capitalists with a
negative debt ratio. This process is illustrated in Figure
\ref{fig:The-formation-of-rentier-tail}. The increase in entropy of
the debt ratio PDF over time is shown.\footnote{Calculated using a per pixel binning of the PDF.}

The greater is the dispersion of the rates of profit\footnote{Technically what we mean here is its coefficient of
  variation, its standard deviation relative to its mean.} and the smaller the gap between the rate of profit and the rate of
interest, the stronger will be this polarisation process and the more
rapid will be the growth in financial entropy.\footnote{As we said earlier the net lending always sums to zero. As such it
is analogous to a motorway with a stream of cars of equal weight heading
past one another at 100kph in opposite directions. The sum of the
momentums of the cars is zero. But if we square their positive or
negative velocities times their weights we get the energy stored in
the traffic. One can view the sum of the squares of the lending positions
of all agents as a bit like energy in that it does not sum to zero
as the activity of the financial system increases but with the important
proviso, that this quantity is not conserved. It can grow without
immediate limit. We will explain later why this growth in financial
energy leads to crises.}

The polarisation process generates a tadpole shape in the phase plane\footnote{It should
be noted that so far this conclusion is based on theoretical arguments and the
evidence from numerical simulations. An empirical investigation of the distribution
of firms on the phase plane remains to be done.}
with a body of active firms clustered round the origin and a rentier
tail of firms that progressively enlongates as the simulation goes
on. Why do we get this asymetrical distribution?

Because the debt ratio of a capitalist can not for long exceed unity.
A capitalist with more debts than assets is technically bankrupt and will
shortly cease trading. There is on the other hand no limit to how
negative the debt ratio of a   capitalist can be. That is
to say, no limit to how much money a capitalist can have in the bank.


What the argument above shows is that it is possible to derive a small
set of laws of motion that characterise the basic development of a
capitalist financial system. The essentials of the phenomenon that has been
called `financialisation', i.e., the growth of financial assets/liabilities relative to real assets,
occurs even in a very simple model. The entropy of the system will increase, polarisation will
occur, and a rentier class will be precipitated out, unless social
work is done to curtail the growth of entropy.

In a more complex model with joint stock firms
a similar polarisation will tend to occur, but in this
case there may be pressure for firms with a very
negative debt ratio (lots of cash) to distribute
this to their shareholders.

This distribution is unlikely to put a halt to the polarisation
process because in the case of cash being distributed from
firms
to individual capitalists the total of positive balances with
the banking system does not change, it just shifts
from abstract legal persons to concrete ones.
The only mechanism by which the polarisation
can be conservatively reduced is by the class of
creditor firms and capitalists to purchase commodities
from the class of debtor firms and capitalists.
This will happen only to the extent that a distribution of money from
cash-rich firms to rentiers results in an increase in the
consumption of the rentier class.

Commodity exchange is a conservative system, i.e., one governed by
conservation laws. This does not apply to taxation. Taxation is a
non-equivalent transfer of value. Heavy taxation of the rentier class
is a form of social work that reduces the entropy of the
system.\footnote{The statistical mechanical effects on money
  distribution under different taxation schemes has recently been
  studied by \cite{Diniz&Mendes2012}.} The introduction of a regulated
or planned economy is an even more powerful form of social work. In a
planned economy the degree of chaos and disorder is reduced and
coherent patterns are established which curb the growth of entropy
characteristic of the free market.

\subsection{Relation to the productive economy}

The model above is very simple, it involves no inter-bank lending
nor the issue of any complex financial instruments. All growth in
debt arises from the behavior of basic actors in the real economy: capitalists and workers. But it is still able to generate the polarisation
of the capitalist class into productive capitalists and financiers.

In a physical system, as it moves from a low entropy state to a high
entropy state we are in principle able extract work from it. Is there
any equivalent in the financial system?

Does it release any analogue of free energy as it evolves?

It is a commonplace observation that an economy with a low initial
level of debt can grow rapidly as the debt builds up, but what is
happening here?

We have said that there is no value absorbed in or contained in the
financial system. The financial system is an information structure
recording the mutual obligations of agents. But this does not mean
that there may not be real value correlates of financial relations.
If we look at all the firms with net debt -- the integral over
the right hand side of the phase plane in Fig.~\ref{fig:phaseplane}  -- then these firms have all
absorbed real resources from the left hand side of the plane. When
capitalists on the bottom half of the phase plane fail to invest and
capitalists on the top half do invest in excess of their income, the
financial system sets up a system of obligations between the debtors
and the saving capitalists. But at the same time, the firms on the
bottom are provided with a market for their output on the top of the
phase plane. There a transfer of real commodities from the bottom
to the top.

A firm on the bottom, produces 10,000 buses containing perhaps 100
million hours of labour which are bought by agents on the top half.
In return the bus company obtains not value but a credit account at
the bank. This is not strictly speaking a commodity exchange. A credit
account in a bank is not value. Had they been paid in gold and got
back 1 million oz of gold bullion, that would have been a commodity
exchange, an exchange of equivalents, since the gold would have contained
real value: the human energy, labour,  required to make it.

The sale of the buses for say 1,000 million Euros is a nonequivalent
transaction in real terms since embodied labour is transfered between
owners without a compensating movement of labour in the other direction.
And, by our previous assumption, it is a transfer between a bus firm
who were deciding to save 1,000 million rather than invest it productively.
This transfer is thus one which would not have occured in the absence
of the credit system. Had gold coins been the medium of exchange\footnote{The existence of gold as a standard of value
is not the same thing as its use as a medium of exchange. The gold standard persisted long
after commodity circulation was generally carried out on credit.},
the million oz of gold would have lain in their safe at the bus factory
and those purchasers who lacked cash (those who are on the right
of the phase plane diagram in Fig.~\ref{fig:phaseplane}) would not have been able to buy the buses.
The sale would thus never have taken place, and the total output of
buses would have had to be 10,000 lower. Buses could still have been
sold to firms who had ready cash, but this would have been a smaller
number.

Readers of Marx's \emph{Capital} \cite{Marx_vol1} may recall
that he analysed the circulation of commodities as being of the form
\[
C\rightarrow M\rightarrow C,
\]
i.e., commodity, money, commodity. He argues that this form already contains
the possibility of crisis.\footnote{``Nothing can be more childish than the dogma, that because every sale
is a purchase, and every purchase a sale, therefore the circulation
of commodities necessarily implies an equilibrium of sales and purchases.
If this means that the number of actual sales is equal to the number
of purchases, it is mere tautology. But its real purport is to prove
that every seller brings his buyer to market with him. Nothing of
the kind. The sale and the purchase constitute one identical act,
an exchange between a commodity-owner and an owner of money, between
two persons as opposed to each other as the two poles of a magnet.
They form two distinct acts, of polar and opposite characters, when
performed by one single person. Hence the identity of sale and purchase
implies that the commodity is useless, if, on being thrown into the
alchemistical retort of circulation, it does not come out again in
the shape of money; if, in other words, it cannot be sold by its owner,
and therefore be bought by the owner of the money. That identity further
implies that the exchange, if it do take place, constitutes a period
of rest, an interval, long or short, in the life of the commodity.
Since the first metamorphosis of a commodity is at once a sale and
a purchase, it is also an independent process in itself. The purchaser
has the commodity, the seller has the money, i.e., a commodity ready
to go into circulation at any time. No one can sell unless some one
else purchases. But no one is forthwith bound to purchase, because
he has just sold. Circulation bursts through all restrictions as to
time, place, and individuals, imposed by direct barter, and this it
effects by splitting up, into the antithesis of a sale and a purchase,
the direct identity that in barter does exist between the alienation
of one's own and the acquisition of some other man's product. To say
that these two independent and antithetical acts have an intrinsic
unity, are essentially one, is the same as to say that this intrinsic
oneness expresses itself in an external antithesis. If the interval
in time between the two complementary phases of the complete metamorphosis
of a commodity become too great, if the split between the sale and
the purchase become too pronounced, the intimate connexion between
them, their oneness, asserts itself by producing \textemdash{} a crisis.
The antithesis, use-value and value; the contradictions that private
labour is bound to manifest itself as direct social labour, that a
particularised concrete kind of labour has to pass for abstract human
labour; the contradiction between the personification of objects and
the representation of persons by things; all these antitheses and
contradictions, which are immanent in commodities, assert themselves,
and develop their modes of motion, in the antithetical phases of the
metamorphosis of a commodity. These modes therefore imply the possibility,
and no more than the possibility, of crises. The conversion of this
mere possibility into a reality is the result of a long series of
relations, that, from our present standpoint of simple circulation,
have as yet no existence.'' \cite[ch.~2, sec.~2]{Marx_vol1}} The potential crisis is caused by the formation of gold hoards,
gold which is withdrawn from circulation and saved. This interruption
of the circuit $C\rightarrow M\rightarrow C$ by money lying idle
means that goods can not be sold: `No one can sell unless some one
else purchases'.

With credit the circuit $C\rightarrow M\rightarrow C$ is replaced
from the standpoint of the seller with one of the form $C\rightarrow F_{a}\rightarrow C$
where $F_{a}$ is a financial asset: a bill of exchange, or a record
of credit with a bank. If both the purchaser and the seller have financial
assets this is the same as the old monetary circulation:
\[
\begin{array}{cccc}
C &  &  & F_{a}\\
 & \searrow & \nearrow\\
 & \nearrow & \searrow\\
F_{a} &  &  & C
\end{array}
\]
The financial asset has just changed places, but if we have purchase
on credit we get:
\[
\begin{array}{cccc}
C &  &  & F_{a}\\
 & \searrow & \nearrow\\
 & \nearrow & \searrow\\
a^{\dagger} & \rightarrow & \rightarrow & C+F_{l}
\end{array}
\]
Here the purchase is funded by the spontaneous creation of a financial
asset/liability pair by the debt creation operation $a^{\dagger}$
giving rise to $F_{a}$ and $F_{l}$. Within this whole process real
value is conserved, since the only real value entering was $C$ and
$C$ leaves having merely changed owner whilst the liability and asset
cancel: $F_{l}+F_{a}=0$.

The `work' done by credit is thus to allow the production and distribution
of goods embodying real labour that could not otherwise occur given
the private organisation of production. This is the free energy extracted
by the increase in entropy of the financial system. Of course the
financial system itself does not do any {\em real} work. The real additional physical work
is done by the employees of companies who are net lenders. The
`work' done by by the credit system is social, it is work to overcome the potential barrier
that private property would otherwise impose on the expansion
of production.

 Commodity exchange requires
and exchange of equivalents. Where these equivalents are not in the
hands of the purchasers, the financial system allows transfers that
are in real terms non-equivalent exchanges, whilst creating instead
a symbolic recompensation for the seller.

This symbolic recompensation in Euros or Dollars is a theoretical command
over future labour. The borrower in this case is the \emph{real appropriator}
of the labour value embodied in the capital goods they acquire. The
lender obtains merely a \emph{formal or symbolic appropriation} of
value.

Private ownership creates a potential barrier to the movement of goods
which credit overcomes. The increasing entropy of the financial system
reflects the `work' that has been extracted in overcoming this barrier
and is why it appears to `create wealth'. It allows the creation of
wealth by labour whose expenditure would otherwise have been inhibited
by the private organisation of the economy.

 If one were to ask a banker
what productive role they paid in the economy the answer would probably
be in terms of the banks providing the finance that the economy
needs.\footnote{``Now let's turn to the purpose of banks in a capitalist
economy. Finance is an intermediary good: You cannot eat it, experience
it, or physically use it. The purpose of finance is to support other
activities in the economy. Banks are meant to allocate capital (funds)
to the best possible use. In a capitalist economy, this means allocating
money to the people or entities that will create the greatest wealth
for the overall society. At the same time, risk management is supposedly
a primary skill for bankers. When capital is allocated well and available
to wealth creating entities, societies flourish. When capital is poorly
allocated, economies can collapse.'' \cite{Judson2012}} Money according
to Adam Smith is the ability to command the labour of
others.\footnote{``Wealth, as Mr Hobbes says, is power. But the person who either acquires,
or succeeds to a great fortune, does not necessarily acquire or succeed
to any political power, either civil or military. His fortune may,
perhaps, afford him the means of acquiring both; but the mere possession
of that fortune does not necessarily convey to him either. The power
which that possession immediately and directly conveys to him, is
the power of purchasing a certain command over all the labour, or
over all the produce of labour which is then in the market. His fortune
is greater or less, precisely in proportion to the extent of this
power, or to the quantity either of other men\textquoteright{}s labour,
or, what is the same thing, of the produce of other men\textquoteright{}s
labour, which it enables him to purchase or command.''\cite[ch.~5]{Smith1974}.} The provision of credit gives a capitalist the authority or permission
to commandeer part of the pool of social labour to his project.

 In
this sense the provision of a line of credit by a bank is like any
other official act of giving permission. It is like for example a building
control office issuing a permit allowing a house to be built. But
the right to hand out such permissions does not make the person handing
them out productive.

 It is bricklayer's and carpenters that actually
produce the house, not the bureaucrat who signs it off. When such
permissions are in demand, the official handing them out may ask for
a cut: ``You want to have this house built, you know your application
might go much more smoothly were you to show your generosity in some way''. In an analogous fashion bankers ask for their cut: interest.

It may appear that a loan to a builder, unlike a building permit, actually gives
the resources to build a house but this is an illusion generated by
the legal relations underwhich it is done. The loan does not give the resources to proceed.
The means are bricks and workers, the
loan gives the right to command these, this is again the removal of a legal impediment
in that a private citizen can not print their own money or issue generally acceptable credits to authorise
the work, whilst banks, unlike other agents, can do this without legal impediment.
The right to command labour is in the end a juridical relation, a more complex and indirect
one than that given by a building warrant, but still ultimately a legal one backed in the end by the
acceptance by the tax authorities  of drafts on the private bank. The fact that the bank
holds an account with the state bank  is what ultimately gives the bank an ability to issue the
authority to command labour. This is in the end a legal delegation of authority by the state.

At one time the charging of interest (usury) as regarded as the
moral equivalent of an official taking a bribe. With the rise of bankers
to political dominance, their very wealth, obtained in this way comes
to be seen as a token of social respectability.\footnote{``This
  disposition to admire, and almost to worship, the rich and the powerful, and to despise, or, at least, to neglect persons of poor
and mean condition, though necessary both to establish and to maintain
the distinction of ranks and the order of society, is, at the same
time, the great and most universal cause of the corruption of our
moral sentiments. That wealth and greatness are often regarded with
the respect and admiration which are due only to wisdom and virtue;
and that the contempt, of which vice and folly are the only proper
objects, is often most unjustly bestowed upon poverty and weakness,
has been the complaint of moralists in all ages.''
\cite[p.~53]{Smith1790_moral}}

\subsection{Contradiction between real and formal appropriation of value}

Why do capitalists seek profits?

On the one hand it is something that is imposed by the nature of competition.
Only profitable firms survive, the more profitable they are the more
likely they are to survive, so a mechanism analogous to natural selection
establishes the profit motive as the driving force of the system.

But there is another way of looking at this.

Whatever its historical form -- gold, banknotes, bank accounts --
money has been the power of command over labour. In all class
societies, members of the upper class had been driven to increase their power over the labouring classes.

Slave owners tried to obtain as many slaves as they could. Feudal
landowners sought to do this indirectly by building up their landed
estates, since attached to the land were serfs. Capitalists do it
by accumulating money. A billion Euros gives a capitalist command
over about 100,000 person years of labour, say 2000 working lifetimes
of indirect labour in commodities, perhaps twice that if they employ
people directly. The more money you have, the more people are at your
command.

Building up a bank balance gives you a symbolic command over lots
of future labour but this is different from really having the product
of this labour, as the fable of Midas long ago pointed out. By the
nature of credit, a thrifty symbolic appropriator depends for his very
existence on the spendthrift debtors.

The credit system socialises this dependence, makes it impersonal,
so that money seems to simply represent abstract value, abstract command
over labour. But the labour it is command over, is that of the debtor
classes as a whole. The sub-prime crisis, like all credit crises, brings
to the fore the fragile and mirage-like quality of this symbolic command,
which can grow beyond any possibility of conversion into a real appropriation.

\begin{figure}[!h]
  \begin{center}
    \includegraphics[width=1.0\columnwidth]{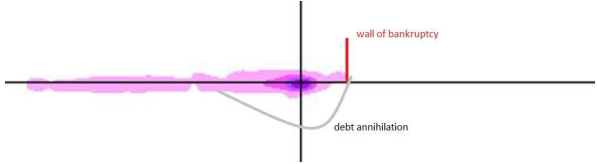}
  \end{center}
  \caption{Capitals pushed against the wall of bankruptcy lead to
    debt/credit annihilation.}
 \label{fig:Capitals-pushed-against}
\end{figure}

Think back to the phase diagram. The symbolic wealth of the rentier
classes is the tadpoles tail. As it wiggles and extends, it drives
the head towards the right towards the wall of bankruptcy (Figure~\ref{fig:Capitals-pushed-against}).
Hit the wall and the credit creation operator $a^{\dagger}$ becomes
its terrible twin the anihilation operator $a$ which cancels out
the debt of the bankrupt at the same time as it wipes out the asset
of the creditor.\footnote{ We admit to `coquetting' with Dirac in our terminology as Marx admitted to doing with Hegel.}

\section{Nonconservation laws of capital accumulation}

If the characteristic or signature of commodity exchange, $C\rightarrow  M
\rightarrow  C$, implies a conservation law, then the signature of
capital,
\[
M \rightarrow  C \rightarrow  M',
\]
 breaks that conservation as
$M' = M + \Delta M$ denotes an increase in the money above the initial
sum spent on commodities in production $M$. Marx famously attempted to
explain how such expansion is possible \cite[ch.~5-6]{Marx_vol1}. It could not, he argued, come
about in the sphere of commodity exchange, since this was governed by a
law of conservation of values. Thus, he argued, profit $\Delta M$
could only be explained outside the realm of commodity exchange, by
the extraction of surplus labour in the production process: Within
capitalist firms -- where Freedom, Equality and Bentham do not
prevail -- the working day is extended beyond the time required to
produce the workers' means of consumption, in order to provide a
surplus that funds profits.

If one takes the aggregate of all capitalist firms, the signature of
this process can be represented as
\begin{equation}
M \rightarrow [ C \Rightarrow (C + \Delta C) ] \rightarrow M',
\label{eq:aggregatecapital}
\end{equation}
where $C \Rightarrow (C + \Delta C)$ represents the production
process that generates a physical surplus product $\Delta C$ after the
consumption of the present working population has been met. Moreover,
through the compulsion of market dependency capitalist firms seek an indefinite expansion
\begin{equation*}
M \rightarrow C \rightarrow M' \rightarrow C' \rightarrow M''
\rightarrow C'' \rightarrow M''' \rightarrow \cdots
\end{equation*}
by reinvesting the profit in greater productive capacity. This process
is therefore governed by nonconservative laws of exponential
growth. Relative growth rates are measured as per unit of time, i.e.,
$[\text{time}^{-1}]$, and a small change in such a rate may produce
dramatic differences over time. For instance, if some quantity grows
at 1\% per annum it doubles in size approximately every 70 years. But
if it grows at 10\% per annum it doubles every 7 years.

As we show below, the nonconservative laws of capital accumulation
interacts with systems outside the capitalist sector which provide it
with labour, natural resources and the material conditions for
habitability in general.

\subsection{Profit rate constraints}\label{sec:profitrateconstraints}

Capitalist firms reinvest their profit -- and rentiers finance
expansion in various branches of production -- based on the rate
of return on the capital invested. Consider a firm $i$ earning annual profit
flow of $P_i$ Euros per annum, before interest, rents and tax
expenditure, with a stock of fixed capital $K_i$ Euros invested in the form
of buildings, machines, equipment etc. The firm then earns an annual
profit rate $R_i = P_i/ K_i$ on its capital stock and the dimension of
this rate is $[\text{time}^{-1}]$. If new investment in the firm is to be
attractive, the profit rate should exceed the real interest rate or
else it would be more profitable to simply earn interest on saved
profits.\footnote{As pointed out by Marx and Keynes \cite{Shaikh2011}.}

\begin{figure}
  \begin{center}
  \includegraphics[width=1.0\columnwidth]{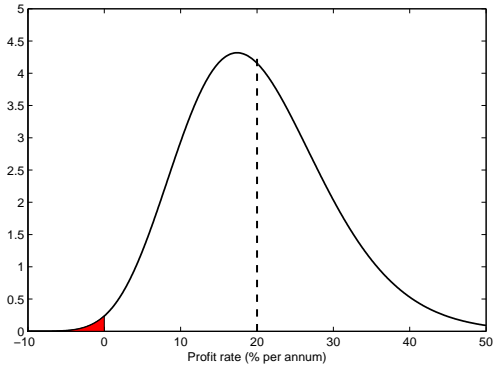}
  \end{center}
 \caption{The distribution of the total capital stock over various
   profit rate brackets. The red-shaded area signifies the proportion
   of the capital that makes a direct loss. The dashed horizontal line
 indicates the profit rate earned when averaged over all capital.}
  \label{fig:profitrate_pdf}
\end{figure}

Next, consider the aggregate of all capitalist firms in the
economy. The total capital stock invested earns different rates of
profit in various firms and branches as illustrated in Figure
\ref{fig:profitrate_pdf}.
Due to varying technical conditions in production and the random
churning of the market, the distribution will always be
dispersed. The process of financial polarisation of capital into net
creditors and debtors, described in the previous section, will also
increase the dispersion of profit rates after net interest
payments. Firms with higher levels of net debt will have lower
retained profit rates, whereas firms accumulating credit accounts
experience higher retained profit rates.

Further, as \eqref{eq:aggregatecapital} signifies, the aggregate
profit flow is predicated on the production of a surplus product of
commodities which requires an annual flow of labour $L$ expended in
production,
\begin{equation}
\begin{split}
M \rightarrow [C &\Rightarrow (C + \Delta C)]  \rightarrow M' \\
& \uparrow\\
& L
\end{split}
\end{equation}

On the basis of the Law of Large Numbers it can then be
shown that average profit rate over all
capital invested, denoted $R$,  is well approximated by
\begin{equation}
R = \rho \frac{ L}{K} \quad \text{(approx.)},
\end{equation}
where $L$ denotes aggregate labour expended per unit of time, $K$
denotes the labour time required to reproduce the total capital stock,
and $\rho$ is the fraction of surplus labour-time performed above the
consumption of the workforce \cite{Farjoun&Machover1983}.

\begin{figure}
  \begin{center}
  \includegraphics[width=1.0\columnwidth]{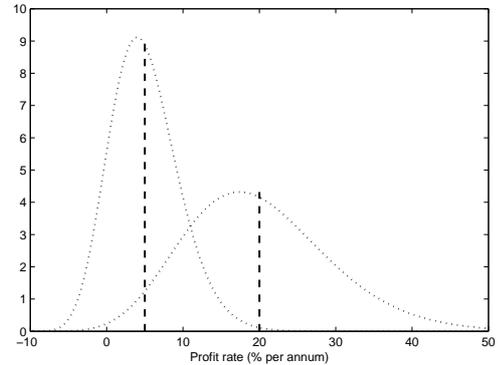}
  \end{center}
 \caption{The distribution of profit rates is deformed as the average
  profit rate $R$ declines from 20\% to 5\% per annum. Note that
the latter distribution is compressed as the proportion of directly
loss-making capital increases and eventually goes bankrupt.}
  \label{fig:profitrate_pdf_shift}
\end{figure}

The significance of the average profit rate $R$ is that it
constrains the entire distribution of profit rates in the capitalist
sector as illustrated in Figure \ref{fig:profitrate_pdf_shift}. Further, it sets the upper limit to the growth rate of the
total capital stock. The average profit rate rises when the amount of
labour expended grows faster than the labour-value of the total
capital stock, and vice versa.

The amount of labour time required to reproduce the capital
stock, $K$, is lowered by productivity growth $g_P$ and the rate of depreciation
$d$, but rises as the surplus product assumes the form of
capital goods invested in production. Based on this insight it is
possible to derive a dynamic equilibrium of the average profit
rate,
\begin{equation}
R^* = \frac{g_L + g_P + d}{\lambda},
\label{eq:R_equilibrium}
\end{equation}
where $g_L$ denotes the relative growth rate of labour expended
and $\lambda$ the ratio of aggregate gross investments to total profits.\footnote{The derivation is based on computing the relative growth $\dot{R}/R = \dot{\rho}/\rho + g_L - \dot{K}/K$, and solving for $R$ when $\dot{R}=0$, cf. \cite{Zachariah2009, CockshottEtAl2008_classical}.} In words, the evolution of the average profit rate is constrained by
the balance between the proportion of
profit that is reinvested, on the one hand, and depreciation,
productivity growth and the growth of labour, on the other.

\begin{figure*}
\begin{center}
 \includegraphics[width=1.0\columnwidth]{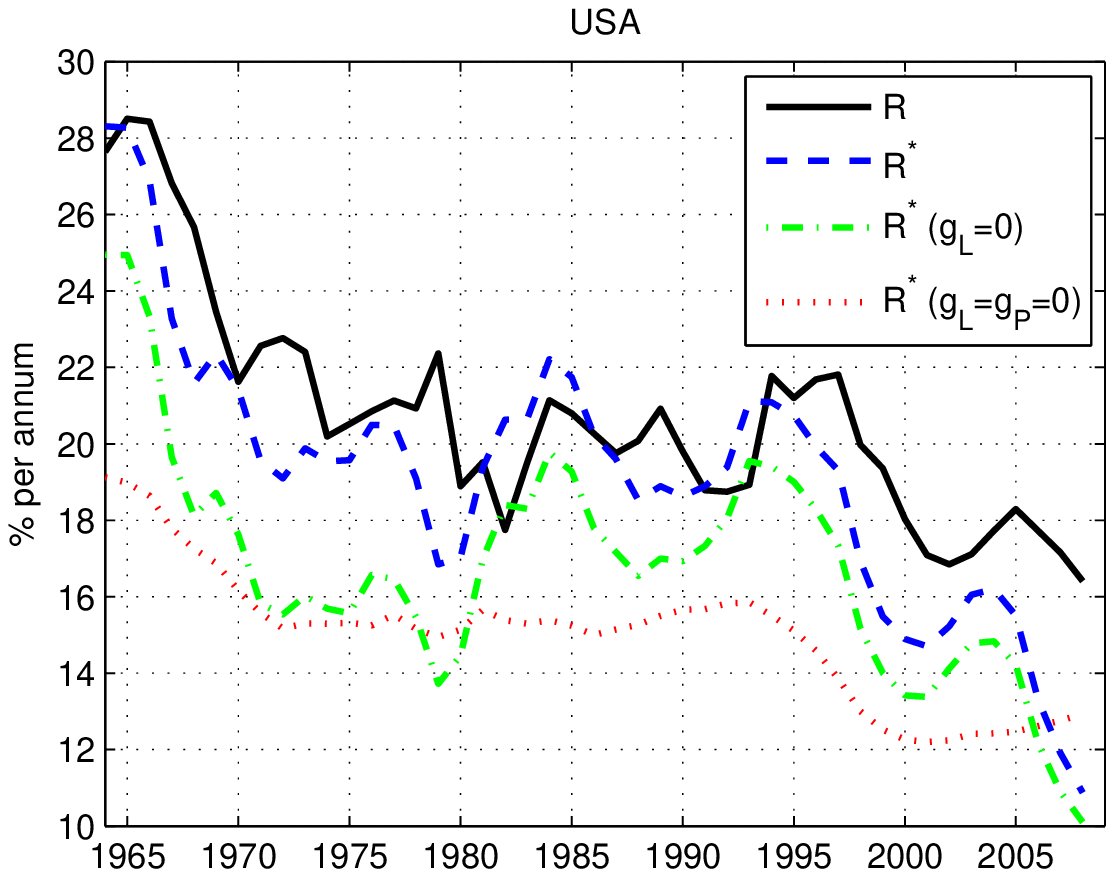}
 \includegraphics[width=1.0\columnwidth]{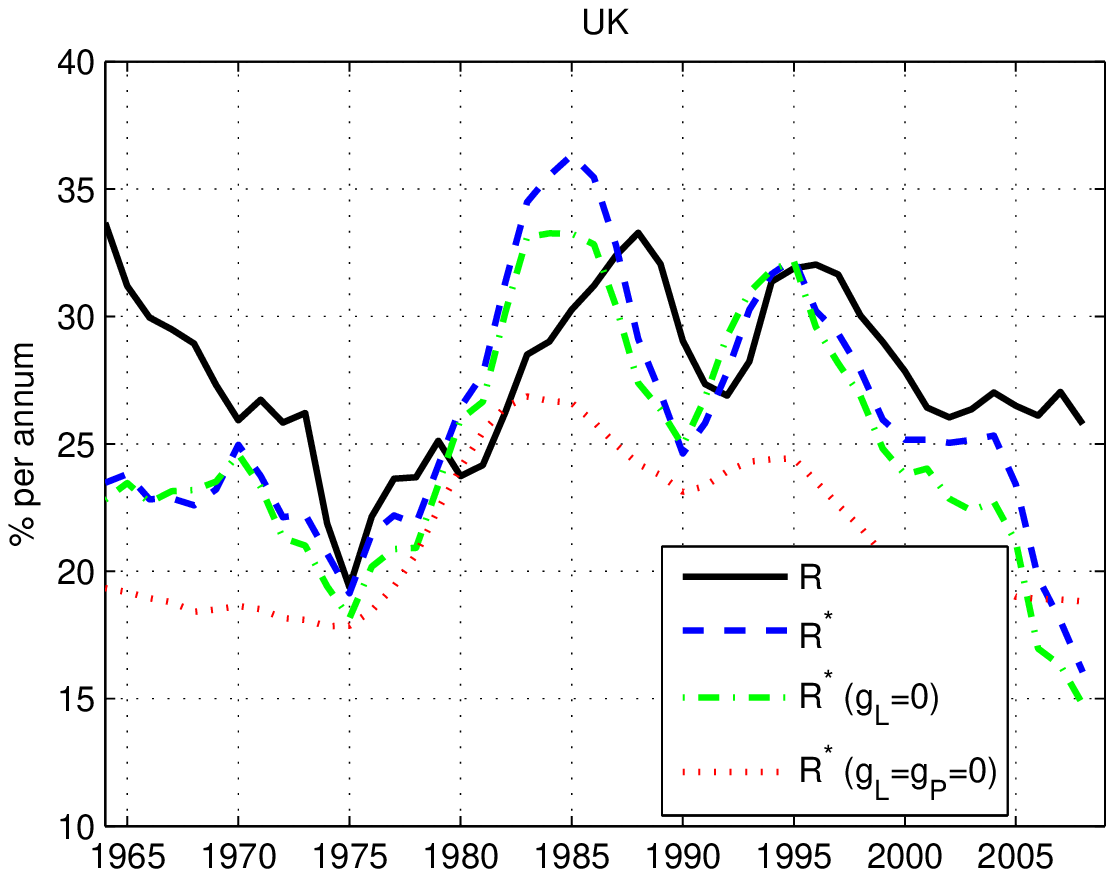}
\caption{Average profit rate $R$ and the equilibrium profit rate $R^*$, USA and UK
    1964-2008. The red dotted curve shows contribution from investments $\lambda$ and depreciation $d$ alone. The green dash-dotted curve shows the additional contribution from productivity growth $g_P$. The blue dashed curve shows the additional contribution from the growth of labour $g_L$. Source: \cite{Marquetti2012} and own calculations.}
\label{fig:profitrate_mean}
\end{center}
\end{figure*}

For sake of illustration, suppose the expenditure of
labour and productivity in the capitalist sector grows by 2\% per
annum and 3\% per annum, respectively. If the aggregate
depreciation rate is 10\% per annum and the ratio of investments to
profits is 0.60, then the average profit rate is inexorably driven to
25\% per annum, irrespective of the distribution between profits and
wages! Figure \ref{fig:profitrate_mean} shows how the trajectory of the average profit rate in the USA and UK have been constrained by the equilibrium profit rate.

\subsection{Resource constraints}

The signature of capital \eqref{eq:aggregatecapital} illustrates that
the capitalist sector is dependent on an element produced outside its
units of production, namely the capacity to work, which provides a
flow of labour, $L$, into the production process. But there are more
such elements. In fact, the process involves an annual flow of any
natural resource $N$ expended in production,
\begin{equation}
\begin{split}
M \rightarrow [C &\Rightarrow (C + \Delta C)]  \rightarrow M' \\
& \uparrow\\
& N
\end{split}
\end{equation}
This includes important inputs such as oil, scarce minerals and arable
land.
Let $g_N$ denote the growth rate of expenditure and $g_\beta$ denote
the rate of regeneration, both relative to the stock of a particular natural resource. Then $g_N - g_\beta$ is the relative depletion rate of the stock
of that resource. Provided the depletion rate is not too rapid,
alternative inputs and technologies can be found; otherwise the
rising scarcity of resource deposits require increasing inputs of
labour in extraction per unit, which lowers productivity of
labour $g_P$. It may also require a higher fraction of profits
reinvested in fixed capital to sustain such
extraction. From \eqref{eq:R_equilibrium} we see that such pattern
of development would lower the equilibrium profit rate.

\subsection{Emission constraints}

Finally, the capitalist sector does not only absorb labour and natural
resources from outside, but produces emissions that deteriorate the
conditions for habitability. For an annual flow of any particular
emission $E$ from production, we can illustrate this as
\begin{equation}
\begin{split}
M \rightarrow [C &\Rightarrow (C + \Delta C)]  \rightarrow M' \\
& \downarrow\\
& E
\end{split}
\end{equation}
The most critical emission is greenhouse gases, in particular
carbondioxide $\text{CO}_2$. Let $g_E$ denote the growth rate
 and $g_\alpha$ denote the rate of absorption of the emission, both relative
 to the stock of accumulated emissions. When the relative emission
 rate, $g_E - g_\alpha$, is positive, the pollution level is rising. This
 acts negatively on the productivity of labour in the agricultural
 sector, which again would lower the equilibrium profit rate \eqref{eq:R_equilibrium}. Furthermore, for a given
living standard, it would necessitate an increasing share of labour
expended in the production of wage-goods consumed by the population as
well as in pollution management. Such a pattern of development would
reduce the relative surplus appropriated by the capitalist classes.

\subsection{Contradictions of exponential accumulation}

In summary, the competitive dynamics arising from capitalist property
relations subject the firms to the compulsion of exponential
accumulation. The capitalist sector as a whole is thus geared towards an
indefinite accumulation of resources -- natural and human -- in the
form of productive assets and commodities. With finite resources the
exponential growth path must continuously hit constraints that need to
be overcome for further accumulation. To an individual rentier such
constraints may appear irrelevant; indeed there is no tangible resource limit
to continuously growing profit income in the form of increasing
symbols in a bank account or on paper bills. These symbols, however,
command real material resources embodied in goods and services.

In the world-historic conjuncture of the twenty-first century each
constraint -- labour $L$, natural resources $N$ and emissions $E$ --
presents obstacles on a scale that cannot be circumvented by the
private capitalist sector alone but would ultimately require massive
state intervention and an increasing public appropriation of the
social surplus.

First, as the length of the working day can only be extended to a
certain limit, exponential growth of labour time expended in the
capitalist sector is ultimately constrained by demographic
factors. Once labour reserves from the countryside have been absorbed,
the growth rate $g_L$ is ultimately set by the rate of population
growth. Further, as societies industrialise the net cost of rearing
children brings the natural population growth rate close to zero, and
is compensated only by net immigration in certain countries. In the
advanced capitalist world as a whole the growth rate of employment was
hovering just below 1\% per annum, prior to the recent crisis. The trend in
East Asia was similar, just above 1\% per
annum.\footnote{The crisis has of course brought  down the growth rate even further, cf. \cite{ILO2012}.} Southern Asia, the Middle East
and Northern Africa will experience the same trend if the regions
follow the path of industrialisation and urbanisation.

The equilibrium profit rate \eqref{eq:R_equilibrium} tells us that
the growth of labour $g_L$ therefore ceases to be a source of
profitability, it is rather the specific balance of investments
$\lambda$ and productivity growth $g_P$ that can sustain conditions
for average profitability in the capitalist sector. This depends on
the specific technological phase of an economy and its institutional
structure of investments. If the technological conditions are not
favourable, the capitalist class is better off dissipating a greater
fraction of the surplus product in the form of luxury consumption
rather than reinvesting profits in the productive capital stock. Only
public enterprises that operate according to break-even criteria,
rather than rate of return on capital investment, will undertake
large-scale investments under such conditions. Short of such
commitments by the state under rentier-dominated capitalism, the
unproductive pattern has sustained mass unemployment in the advance
capitalist countries since the 1980s \cite{Stockhammer&Klar2011}.

Second, the competitive and anarchic dynamics of the capitalist
sector inexorably drive firms to exploit energy sources and
materials that can satisfy the compulsion of exponential
accumulation; if only over a short time horizon. The early capitalist
factory system that arose in England was initially powered by
water. But the supply of labour was concentrated in populous towns;
hence factories that employed more mobile steam-engines could
establish there and get a competive edge. Unlike the water wheel,
steam-engines were powered by a scarce fossil fuel: coal. This was the
genesis of fossil fuels as the primary source of
energy in capitalism \cite{Malm2013}. Further,
exponential accumulation must be met by exponential consumption for
profits to be realised. This does not only increase the scale of
material inputs but also reduces arable land and fresh water supplies.

Misallocation of critical scarce resources through the market
mechanism can thus only be counteracted by state intervention to
bring down the depletion rates to viable levels, i.e., imposing
quotas and a controlled scale of extraction. Further the composition
of consumption among higher-income populations would need to be steered away from material goods and towards sustainable
services, including socialised forms of consumption. Public taxation
as well as lower workhours provide mechanisms for changing
consumption patterns.

Third, emissions of greenhouse gases impose serious risks of
deteriorating the viability of the global capitalist economy. Under
current global emission trends the average world temperature will
increase beyond +2 degrees Celsius by the year 2050, and rapidly
approaching +4$^\circ$C. The effects of such climatic changes are
an increased frequency of heat waves; decline in crop yields;
exacerbated water scarcity; high-intensity tropical cyclones; and
irreversible loss of biodiversity with unpredictable effects on human
life \cite{WorldBank2012}. All of which undermine the
reproduction of the source of value -- the workforce.

The bulk of emissions in the global economy, about one quarter, arises
in electricity and heat, followed by industry, transport and
agriculture. Deforestation further exacerbates this by reducing the
absorption rate \cite{Herzog2009}. Reducing per capita
emissions at a sufficient rate to counteract global warming requires
large-scale investments in power generation and an energy-efficient transformation of the assets in
production, transport and housing. This is well beyond the what
private research and development in the capitalist sector can commit
to alone. As the material conditions deteriorate, a coordinated effort
by states akin to wartime planning will become increasingly necessary
to sustain a viable productive sector and workforce.

\section{Contradictions between laws of market exchange and capital accumulation}

We now consider the dynamics that arise from the interaction between the conservative laws of market exchange and the nonconservative laws of capital accumulation.

\subsection{Value conservation and the signature of capital}

 Marx   only partly answered the problem of ``where the
money comes from'' in the process $M\rightarrow C \rightarrow M'= M + \Delta M$. He explained how capitalists obtained a net income
from their capital, but this was only half the problem. If the capitalists
follow the maxim ascribed to them by Marx -- ``Accumulate, accumulate!
That is Moses and the prophets!''  -- then the signature of capital
$M\rightarrow C\rightarrow M'$ extends into
\[
M\rightarrow C\rightarrow M'\rightarrow C'\rightarrow M''\rightarrow
C''\rightarrow M'''\rightarrow \cdots
\]
 which requires exponential growth in the quantity of money. In the
19th century, the British economy, like most others, depended on precious
metal for its monetary base. An exponential growth in the quantity
of money implies the same sort of growth for gold stock. But if we
look at historical data for the growth of the world gold stock, we
find that during the 19th century it was growing at well under 1\%
per annum. Given that the British economy grew at over 2\% a year,
there was a discrepancy between the growth of gold and the growth
of commodity circulation.

\begin{table}[h]
\label{tab:gold} %
\begin{center}
\caption{Growth of the world gold stock, 1840 to 2000. Reproduced from \cite{CockshottEtAl2008_classical}}
\begin{tabular}{lrc}
 &
\multicolumn{1}{c}{Stock} &
Annual growth \tabularnewline
\multicolumn{1}{c}{Period} &
\multicolumn{1}{c}{(million troy oz.)} &
(percent) \tabularnewline \hline
1840--1850  &
617.9 \hspace{2em}  &
$0.27$ \tabularnewline
1851--1875  &
771.9 \hspace{2em}  &
$0.89$ \tabularnewline
1876--1900  &
953.9 \hspace{2em}  &
$0.85$ \tabularnewline
1901--1925  &
1430.9 \hspace{2em}  &
$1.64$ \tabularnewline
1926--1950  &
2130.9 \hspace{2em}  &
$1.61$ \tabularnewline
1951--1975  &
3115.9 \hspace{2em}  &
$1.53$ \tabularnewline
1976--2000  &
4569.9 \hspace{2em}  &
$1.54$ \tabularnewline
 &
 &
\tabularnewline
\end{tabular}
\end{center}
\end{table}

Since gold stocks could not grow fast enough to support the expansion
of the economy, capitalists had to resort to commercial bills. An
Iron Master taking delivery of coal would typically write a bill of
exchange, a private certificate of debt, promising to pay within 30
or 90 days.

Payment of wages would generally have to be done in cash. Capitalists
have tried at times to pay wages in tokens redeemable only at company
stores (`scrip') but legislation by the state, eager to maintain
its monopoly of coinage if not to defend the interests if the workers,
tended to put a stop to this. Payment in cash represents a transfer
from the safes of capitalists to the pockets of their employees, with
a corresponding cancellation of wage debts. At the end of the week,
the wage debt has been cleared to zero, and there has been an equal
and compensating movement of cash.

Workers then spend their wages on consumer goods. For the sake of
simplicity, assume that there is no net saving by workers so that
in the course of the week all of the money they have been paid is
spent. This implies that immediately after pay-day, the money holdings
of the workers are equal to one week's wages. If these wages were
paid in coin this would have set a lower limit to the quantity of
coin required for the economy to function.

When workers spend their wages on consumer goods they transfer money
only to those firms who sell consumer goods -- shopkeepers,
inn-keepers and so on. We can expect these firms not only to make up
the money they had paid out in wages, but to retain a considerable
surplus. The final sellers of consumer goods will thus end up with
more money than they paid out in wages. From this extra cash they can
afford to redeem the bills of exchange that they issued to their suppliers.

In the absence of bank credit, suppliers of manufactured consumer
goods would be entirely dependent for cash on money arriving when
the bills of exchange, in which they had initially been paid, were
eventually redeemed by shopkeepers and merchants. The payment situation
facing raw materials firms was even more indirect: they could not
be paid unless the manufacturers had sufficient cash to redeem bills
of exchange issued for yarn, coal, grain, etc.

The process of trade between capitalists leads to the build-up of
inter-firm debt. We suggest that the total volume of inter-firm debt
that could be stably supported would have been some multiple of the
coinage available, after allowing for that required to pay wages.
If one takes the aggregate of all firms the ideal signature of this
process can be represented as:
\[
M\rightarrow[C\Rightarrow(C+\Delta C)]\rightarrow M+\Delta M
\]
 where $[C\Rightarrow(C+\Delta C)]$ represents the production process
that generates a physical surplus of commodities after the consumption
needs of the present working population has been met. If there is
no new issue of coin by the state then the $\Delta M$ cannot be `real
money'; rather, it must be in the form of bills of exchange and
other inter-firm credit.

For the capitalist class considered as a whole this should not be
a problem since the $\Delta M$ is secured against the accumulated
commodity surplus $\Delta C$. There is a net accumulation of value
as commodities, and accounting practice allows both the debts owed
to a firm and stocks of commodities on hand to be included in the
value of its notional capital. As the process of accumulation proceeds
in this way the ratio of commercial debt to real money will rise.
If the period for which commercial credit is extended remains fixed
-- say at 90 days -- then a growing number of debts will be falling due each
day. If these have to be paid off in money, then a growing number
of firms will have difficulty meeting their debts in cash.

The basic contradiction between capital's exponentially growing need
for money and the much slower growth of gold production led to a series
of transformations of the monetary system during the 19th and 20th
centuries.\footnote{``One of the principal costs of circulation is
  money itself, being value in itself. It is economised through credit in three ways. A. By dropping away entirely in a great many transactions.
B. By the accelerated circulation of the circulating medium. ... On
the one hand, the acceleration is technical; i.e., with the same magnitude
and number of actual turnovers of commodities for consumption, a smaller
quantity of money or money tokens performs the same service. This
is bound up with the technique of banking. On the other hand, credit
accelerates the velocity of the metamorphoses of commodities and thereby
the velocity of money circulation.
C. Substitution of paper for gold money.'' \cite[ch.~27]{Marx_vol3}}
\begin{enumerate}
\item Gold was supplemented by commercial credit. This displaced gold from
transactions between capitalists.
\item Commercial credit was supplemented by the discounting of bills of
exchange by the banks.
\item Payment in bills of exchange was largely replaced by payments by cheque
and commercial credit by bank credit.
\item Gold coins were withdrawn from circulation to be replaced by banknotes
with gold only used in settlements between international banks. This
meant that wage payments no longer depended on precious metal.
\item National currencies were then completely removed from the gold standard,
and state notes became the base money. This was completed by the withdrawal
of the dollar from the gold standard in the 1970s.
\item With the development of computerisation in the third quarter of the
20th century it became practical to pay wages directly into bank accounts.
This meant that state base money circulating could be substantially
less than the monthly or weekly wage bill.
\item Finally with the general issue of credit cards, the credit system
spread from the capitalist class to all classes in society.
\end{enumerate}
When money was still gold, this gold was value -- it was embodied labour
and had a value internationally because in all countries the production
of gold required a great deal of labour. With the modern system of
national and supernational currencies money is no longer value. No
significant work goes into the printing of  100 Euro notes.
Why then can they function the same way that gold used to?

Central bank notes used to be issued under the gold standard, but
these were just tokens for gold, and could be redemed for bullion
on demand at the central bank. There is no promise by the ECB to redeem
Euros to gold at any fixed exchange rate.\footnote{Though collectively the central banks of the Euro area held
  over 430 billion Euros in gold reserves. These can potentially be used in settlement
with other central banks to settle foreign trade debts, but in practice
there is little or no intervention by the ECB using gold to support
the value of the Euro.} Since there is no definite link between the Euro and gold or between
the Pound and gold, how can these currencies function as a measure
of value and medium of exchange?

Why are they worth anything?

Well for one thing they are legal tender, but what does this mean
and why is it important?

Throughout the Eurozone the following is held to apply in cases where
a payment obligation exists:
\begin{quote}
\textbullet{} Mandatory acceptance of Euro cash; a means of payment
with legal tender status cannot be refused by the creditor of a payment
obligation, unless the parties have agreed on other means of payment

\textbullet{} Acceptance at full face value; the monetary value of
a means of payment with legal tender status is equal to the amount
indicated on the means of payment

\textbullet{} Power to discharge from payment obligations; a debtor
can discharge himself from a payment obligation by transferring a
means of payment with legal tender status to the creditor. \cite{ELTEG2010}
\end{quote}
The circulation of the Euro is legally enforced in the relationship
between shops and customers, but between businesses they can in principle
both agree to settle obligations in something other than Euros. But
with retail commerce and all taxes payable in Euros, its circulation
is effectively enforced.

But there is a difference between the previous generation of state
monies like the Franc or D-mark and the Euro. The previous generation
fell into the general category of state token monies. This is a very
old category of money \cite{Ingham2004,Knapp1973,Wray2004}.
Precious metal money was prevalent
in early modern Europe. But in China the monetary system was from
a much earlier stage based either on copper tokens or on paper notes \cite{Glahn1996,Glahn2004}
and it is arguable that for much of the Roman Empire the Denarius
was little more than a copper token with merely a symbolic coating
of silver \cite{Bolin1958}. In such a system the circulation of money
goes illustrated in table \ref{tab:circulation}.
\begin{table*}
\caption{Circulation of money}
\label{tab:circulation}
\begin{center}
\begin{tabular}{cccccccc}
\emph{\small State} &
 &
\emph{\small Pay} &
\emph{\small Lackeys} &
\emph{\small Purchases} &
\emph{\small Producers} &
\emph{\small Tax} &
\emph{\small State}\tabularnewline
{\small Emperor/} &
 &
 &
 &
 &
 &
 &
{\small Emperor/}\tabularnewline
{\small{} King} &
 &
{\small $\rightarrow$} &
{\small Soldiers} &
{\small $\rightarrow$} &
{\small Artisans} &
{\small $\rightarrow$} &
{\small King}\tabularnewline
{\small $\uparrow$} &
 &
 &
{\small Bureacrats} &
 &
{\small Peasants} &
 &
{\small $\downarrow$}\tabularnewline
{\small $\nwarrow$} &
 &
 &
{\small $\leftarrow$} &
 &
{\small $\leftarrow$} &
 &
{\small $\swarrow$}\tabularnewline
\end{tabular}
\end{center}
\end{table*}

The empire or state imposes the circulation of its
token currency by obliging the producers to pay taxes in money. Since
the producers must `render unto Ceasar', they are forced to sell their
product to the employees of Ceasar. The state creates money tokens
with which it pays its employees. The state employees willingly work
for the state in return for these tokens knowing that these tokens
will enable them to command the labour of others in their turn. The
state thus breaks down the self-sufficient or barter economies of
the countryside and enforces the spread of commodity exchange. Forestater
gives a dramatic account of how this process was enforced in the British
Empire \cite{Forstater2003}.

In a pre-monetary tax system, the tax is levied in the form of a direct
duty on the population to work for or deliver goods to the state.
In this case the real appropriation of labour by the state is directly
visible. In a monetary tax system with token money the \emph{real
appropriation} of surplus labour occurs when soldiers and other state
employees deliver goods and services to the Emperor. There is a distinct
\emph{formal or symbolic appropriation} when the taxes are levied
on the producers. This is similar to the relationship that we previously
analysed between formal and symbolic appropriation in the credit system,
but with this difference: the formal transfer in the tax system has
no pretence of equivalence.

The value of a state token money is based on something historically
prior to commodity production, something that goes right back to the
earliest state forms: the power of the state to command the labour
of its inhabitants. The value of the Swedish Krona is set by the fact
that the Swedish state (Crown) directly or indirectly appropriates
more than half the labour in the Kingdom.\footnote{This is a slight
  simplification, public expenditure is over 53\% of GNP, but a part
  of that is indirect appropriation of labour as social protection,
  where individual citizens are paid social benefits in Krona.} The sum in Krona paid out for that directly social labour has its
value set by this labour that the Crown directly commands. This rate
of exchange between royal tokens and labour sets the monetary equivalent
of labour time in Sweden, which then operates via the medium of commodity
trade within the remaining private part of the economy. The direct
royal command over the labour of his subjects is then symbolically
appropriated by the capitalist class in the form of Krona credits
in the Svenska Handelsbanken, etc. to give them command over the privately
employed working class.

For a monetary system like this to work, you need a clearly defined
and controlled territory of the empire, an efficient tax system, and
a state that commands a substantial portion of the total labour of
society. And that state has to have the sovereign right to issue its
own currency. These conditions are not met in the Eurozone.

\subsection{Formation of interest}

The formation of credit and debt relations in the capitalist sector
demands further an analysis of the formation of interest in the
banking system. The specific feature of this system is deposit taking.

A deposit-taking banker is in a perilous position since
he accepts cash over and above his own capital which he then lends
out. Because he has lent out cash that was not his own capital, and
is under an obligation to encash deposits on demand (or after some
fixed warning period), he can easily become insolvent. The remaining
cash he holds in his safe is never enough to meet his maximum obligation
to his creditors. Should the day dawn on which too many of them demand
their money back, he is lost.

Suppose we model this as a stream of customers arriving at the banker's
till at random intervals. Each customer either makes a deposit or
a withdrawal. The customers may make a withdrawal of any amount up
to their current credit balance. We further assume that in a steady
state customers are as likely to make a deposit as to make a withdrawal.
Then the more customers that a bank has, the smaller will be the proportional
variation in the withdrawals from day to day.

As the number of customers rises, the variation in the amount withdrawn
in any week falls, and so too does the maximum withdrawal that can
be expected. A very small bank would have to keep all its deposits
in the safe as an insurance against having to pay them out, but a
bank with 20,000 customers might never see more than a few percent
of its cash deposits withdrawn in any week. A bank with that number
of customers could safely issue as loans several times as much in
paper banknotes as the coin that it held in its vaults, safe in the
knowledge that the probability of it ever having to pay out that much
in one day was vanishingly small. Thus with a starting capital of
10 million the bank could lend out 200 million after building up its
network of customers. This creation of new paper money by the banks
was the hidden secret behind the signature of capital.

The cost to a bank of making a loan is related to the likelihood that
the reserves left after the loan will be too small to cover fluctuating
withdrawals. If this happens the bank may lose its capital.

\begin{figure}
  \begin{center}
    \includegraphics[width=1.0\columnwidth]{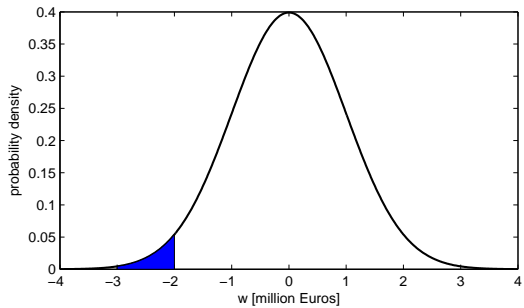}
  \end{center}
  \caption{Example PDF of maximum excursion $W$ of a bank's reserves during a year. Shaded area gives the probability of withdrawals $W$ between 2 and 3 million Euros.}
\label{fig:withdrawals}
\end{figure}

Let $W$ denote the maximal excursion of reserves from their mean
position during a year, due to random deposits and withdrawals by customers. Let us assume that $W$ follows a Gaussian distribution, with a standard deviation of 1 million Euros.
 Suppose that the banker had a capital of 5,000,000
and that he would lose his capital if the bank failed. Then, if he
started with reserves of 3 million, making a loan of 1 million would
reduce the reserves to 2 million and the loan would have an expected
cost of $5,000,000 \times P_{e}$, where $P_{e}$ is the probability
that the withdrawals $W$ exceed the resulting reserve level \emph{after} the
loan.\footnote{In other words, $P_e = \Pr\{ -3 \text{ million} < W \leq -2 \text{ million} \} \approx 0.022$ for a zero-mean Gaussian variable $W$ with standard deviation $\sigma = 1 \text{ million}$.}  Figure \ref{fig:withdrawals} illustrates this.
This amounts to an expected cost of about 107,000, which sets a
lower limit on the interest it would be rational for the banker to charge
for the loan, namely 10.7\% in this case.

Different banks will charge different rates of interest, but through
the pressures of bankruptcies the lower safety rate is likely to emerge
from capitalist banking practices. As the ratio of reserves to deposits
fall, this pressure would be reflected in higher interest rates. There
would thus be an inverse relationship between the reserve to deposit
ratio and the interest rate.


At the birth of the banking system the reserves of the private banks
were in bullion. Now the reserves are in the form of credit accounts
with the state bank. These come into existence when the state
pays people for services rendered, or makes state grants. One of the
authors recalls that his student grant, and payments for attending
Research Council meetings were in the form of bits of paper that
looked like cheques but instead of being drawn on a bank, they
were drafts made out on `The Queen and Lord Treasurer's Remembrancer'.
Private banks can present these government drafts to the Bank
of England, and have their account with the Bank of England
credited by a like amount.

The other side of the coin is when a citizen pays taxes to the
Exchequer using a cheque drawn on a commercial Bank. The
Exchequer then passes these to the Bank of England which writes
down the credit account of the relevant commercial bank.

If $B$ are the reserves of the private banking system, it follows that
$\frac{dB}{dt}=G -T - S$ where $G$ is government payments, $T$ are  tax
payments and $S$ is the sale of government securities, all per unit time. Sales of government securities obviously reduce the monetary reserves of the private banking system.

By adjusting its sales of securities, the state can manipulate the
reserves of the private banking system and thus control the rate
of interest.  In a state with its own central bank, it is an illusion to think
that the government is at the mercy of the `money market' when
selling these securities. A state where $G>T$, i.e., one with a budget
deficit can if it wishes simply refrain from selling securities for a while
and in due course the growth of the reserves of the private banks
will force a fall in the market interest rate. In the end, the private banks are forced to buy government debt. Any  reserves
exceeding the level required for the safety of their loans to the
private sector are simply dead assets, yielding no income. The only way for the banking system as a whole to get income from these
assets is to buy government securities.

 If the state bank wishes,
it can  buy back government securities from the private banks
forcing the interest rate down even faster. This is the mechanism
by which so-called Quantitative Easing reduces the interest rate.

\subsection{Dynamics of sector balances}

\begin{figure*}
\begin{center}
 \includegraphics[width=1.0\columnwidth]{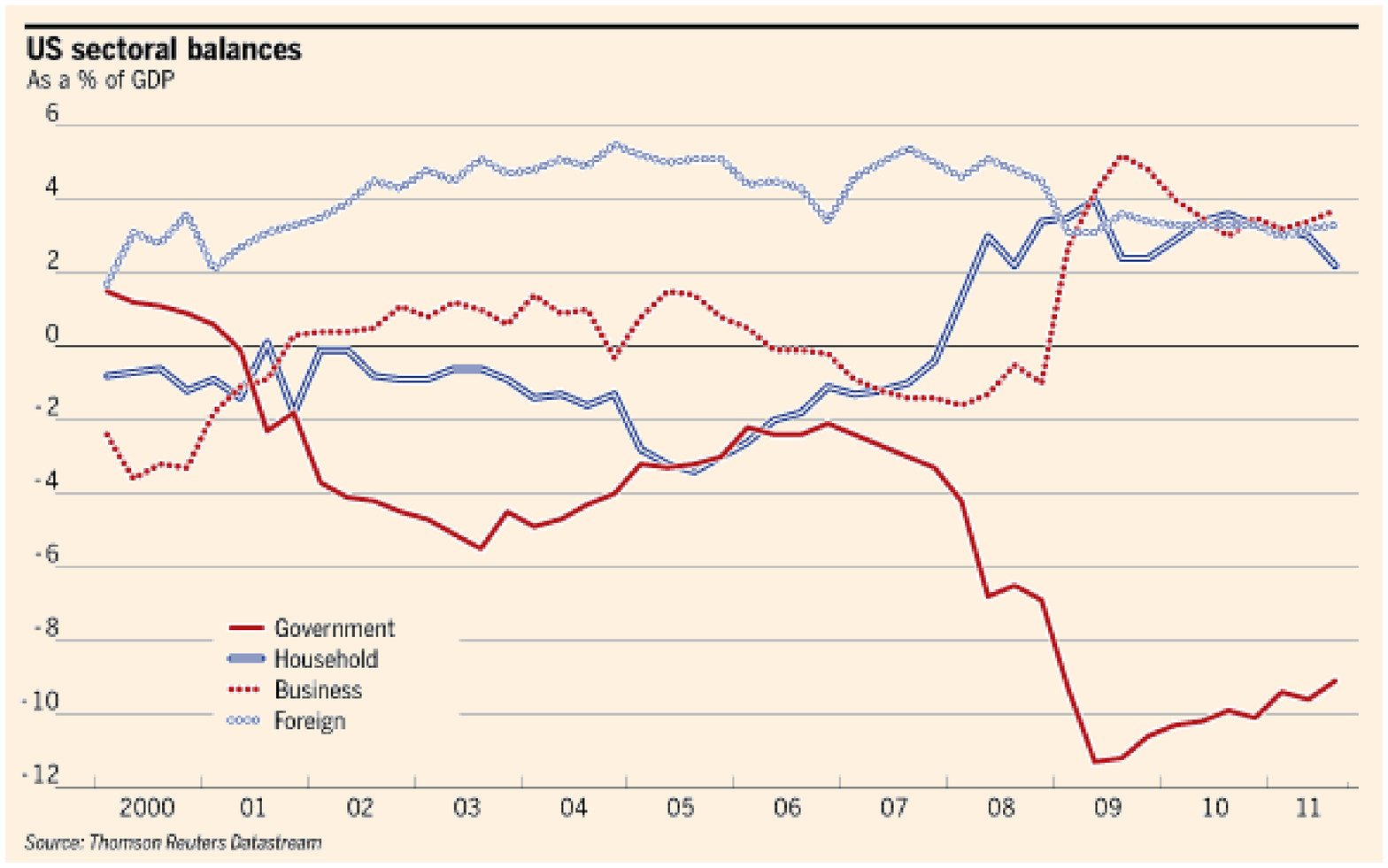}
 \includegraphics[width=1.0\columnwidth]{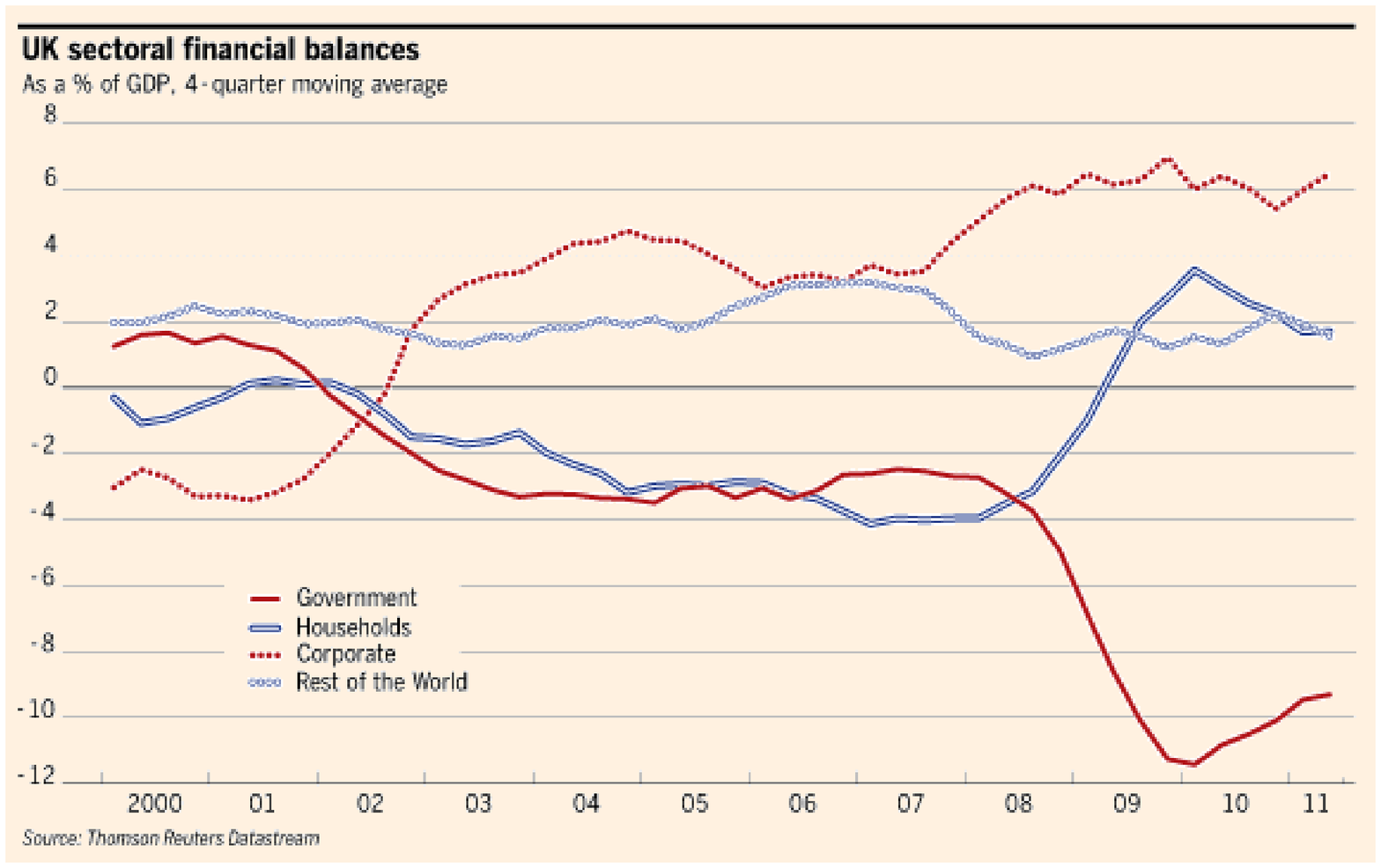}
\caption{Trends in sectoral balances in US and UK. Reproduced from http://www.concertedaction.com/wp-content/uploads/2011/12/US-Financial-Balances.png and .../UK-Financial-Balances.png }
\label{fig:balance}
\end{center}
\end{figure*}

The reader may recall equation \eqref{eq:conservationofmomentum}
showed that the sum of borrowing and lending have to balance. That
equation was defined over all agents, but it also applies if,
instead of agents, we consider economic sectors:
households, non-financial companies, banks, the state, and
the `rest of the world'. Figure \ref{fig:balance} illustrates these
sector balances for the USA and UK. In Table
\ref{tab:Sectoral-balances-for} we present such data for the Eurozone.
  Now the sum of the sectoral balances have to equal 0. Look at the business sector  you can see that, as for most years, this
 sector was running a financial surplus. This surplus was only
 possible because of the deficits, i.e., borrowing, by the household
 and  state sectors. After the sub-prime mortgage crisis lending to
 working-class and middle-class borrowers became much more
 restrictive and the household sector became net lenders. This does not mean that the working class became net lenders of course, it
 is just that the propertied and working classes are aggregated in the
 household sector statistics. Once working-class borrowing fell, then
 the saving by the propertied classes became the dominant factor in the household sector. This constrains the market for consumer goods producing the recession.

 The balances make it clear that the business sector can only run a
 financial surplus
 if other sectors run a deficit, and given the poor competitive
 position of US and UK manufacturing
 the rest of the world is not going to run a deficit with them. That
 leaves only the
 household or the state as possible absorbers of the financial surplus
 of the corporate sector.

 The limitation of the financial system to support borrowing in 2008 had in the short term little to do with profitability of industry, as it was lending to working class households not to industry. Their ability to
 borrow was constrained by low wages. The repackaging of mortgages hid
 the fact that a large portion of working class borrowers were bound to default on their loans, but it could only hide it for a while,
 eventually the  poverty of the working class borrowers became the
 determining factor. In longer term though one has to ask why the
 corporate sector ran a financial surplus? In other words, saving a
 portion of its profit income rather than reinvesting?

\begin{figure*}
\begin{center}
 \includegraphics[width=1.0\columnwidth]{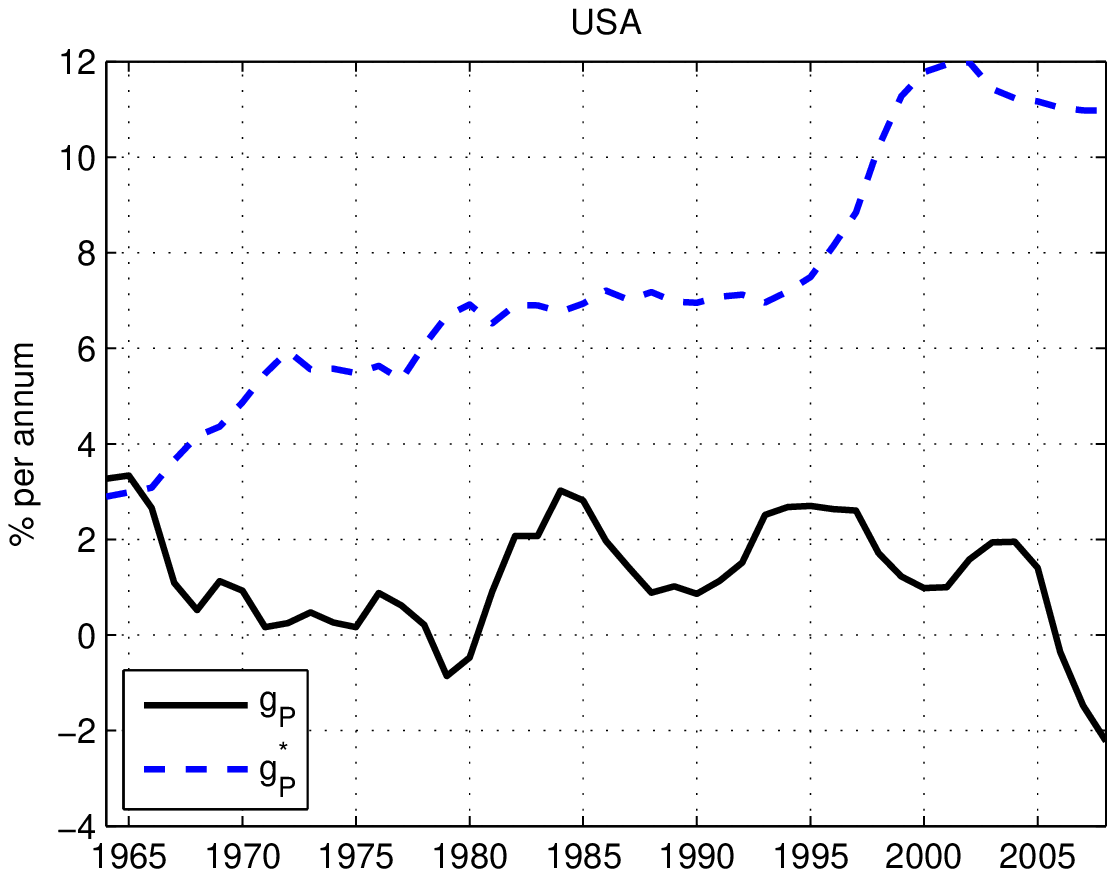}
 \includegraphics[width=1.0\columnwidth]{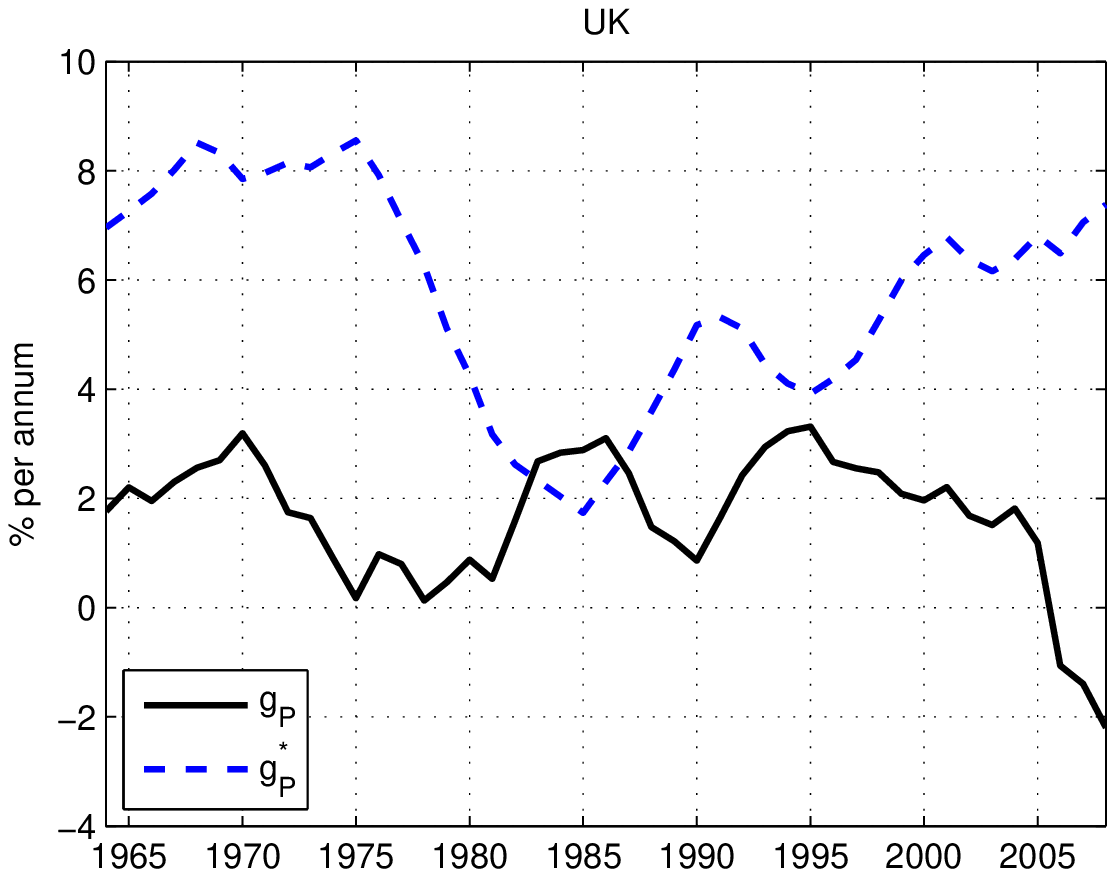}
\caption{Actual net productivity growth $g_P$ and the growth $g^*_P$ required to sustain average profitability at the level in 1964, under actual trends of demographics and investments. Using \eqref{eq:R_equilibrium}, $g^*_P =\lambda R_{\text{1964}} - (g_L + d)$.}
\label{fig:productivitygrowth}
\end{center}
\end{figure*}

The share of profits expended as net investments has rarely exceeded 25\% in the advanced economies, since the post-WWII boom \cite{Marquetti2012}. We would argue that this is a structural feature of a modern capitalist economy. As the demographic transition is completed and
production technologies mature, the growth rates of labour, $g_L$, and
productivity, $g_P$, become moderated. Then it becomes increasingly
difficult to maintain average profitability when firms are reinvesting
a large portion of their retained profits, as seen in equation
\eqref{eq:R_equilibrium}. Fig.~\ref{fig:productivitygrowth} illustrates the productivity growth required to sustain average profitability at the post-WWII boom levels, under the actual demographic and investment trends. Short of such such spectacular growth rates, rapid capital accumulation is unsustainable in the capitalist sector and a substantial portion of profits will instead be accumulated as financial surpluses and spent on luxury consumption.

\section{The crisis in the Eurozone}

From the standpoint of exchange being a conservative system, the exponential
growth of value implied by compound interest has to be a temporary
\emph{disequilibrium} phenomenon -- tied to an exponential growth in
whatever is the source of value. If the expenditure of human energy
is the source, then the accumulation of capital value must depend
on a similar exponential growth of the working population.

\begin{figure*}
     \begin{center}
        \subfigure{%
            \includegraphics[width=1.0\columnwidth]{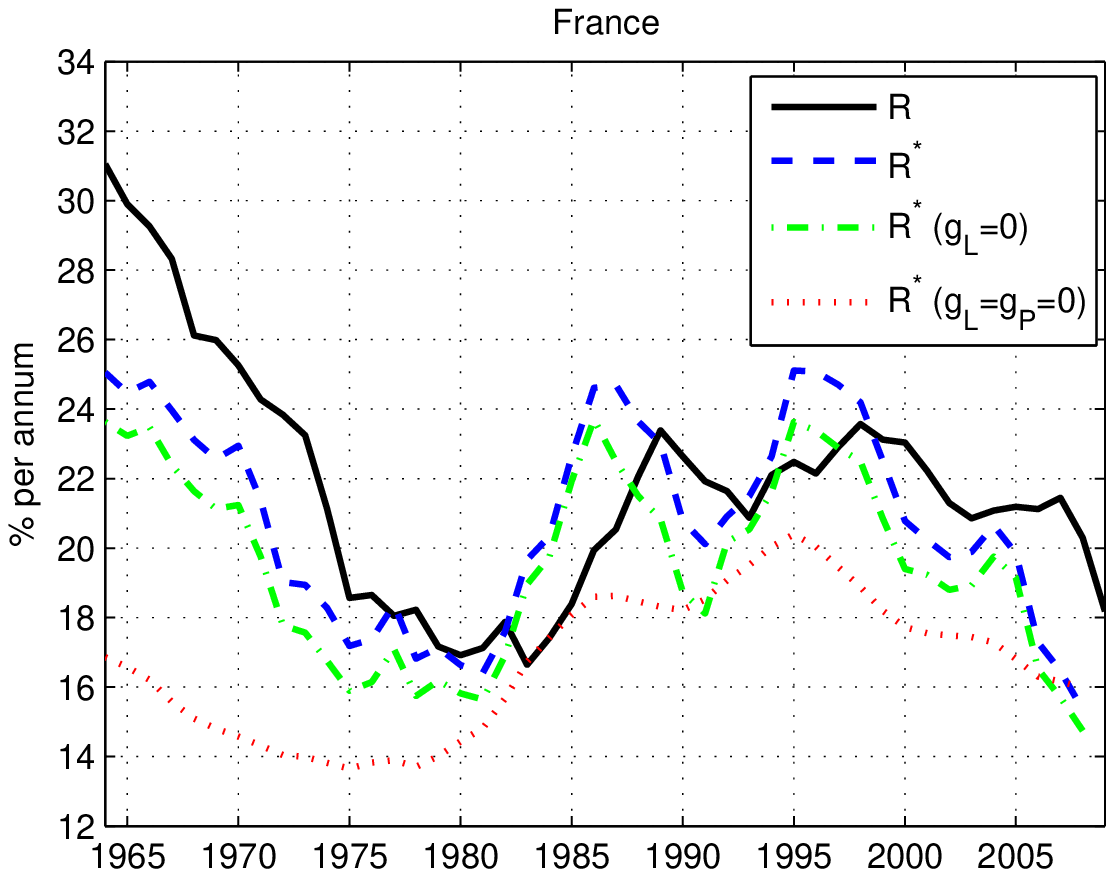}
        }%
        \subfigure{%
           \includegraphics[width=1.0\columnwidth]{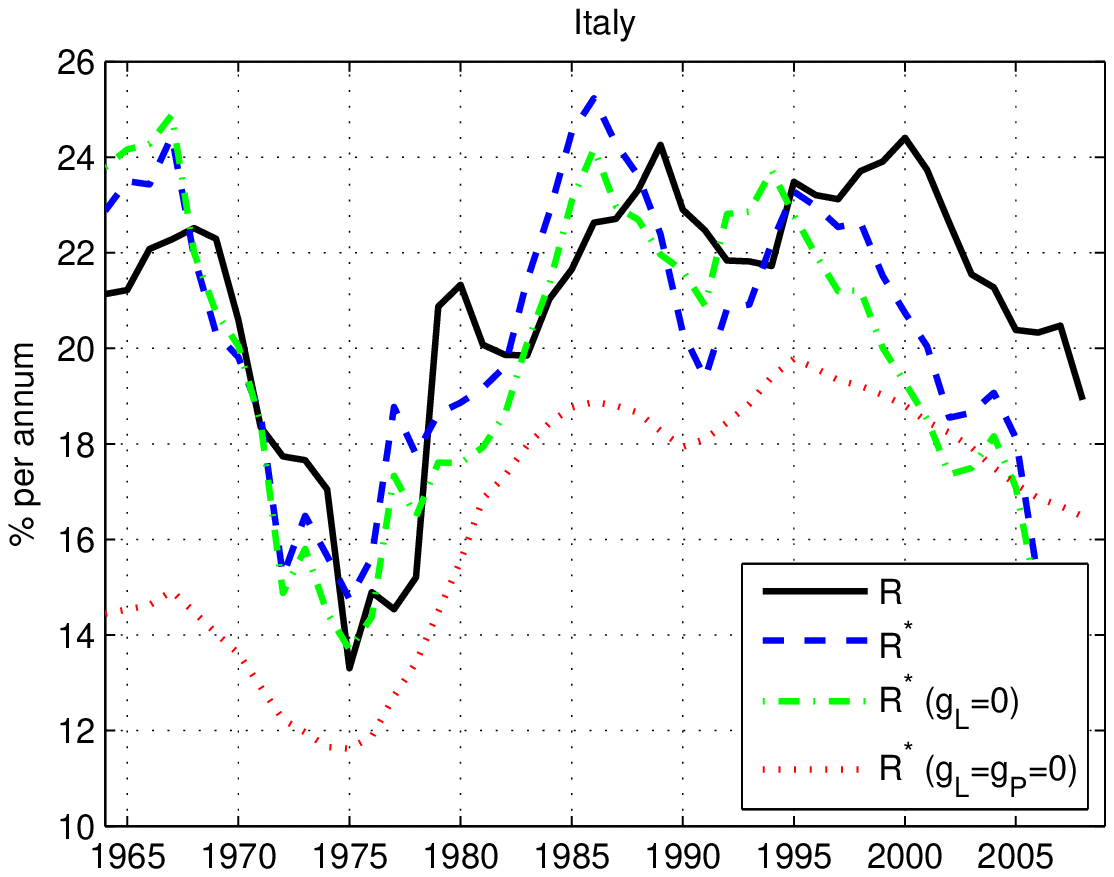}
        }\\ 
        \subfigure{%
            \includegraphics[width=1.0\columnwidth]{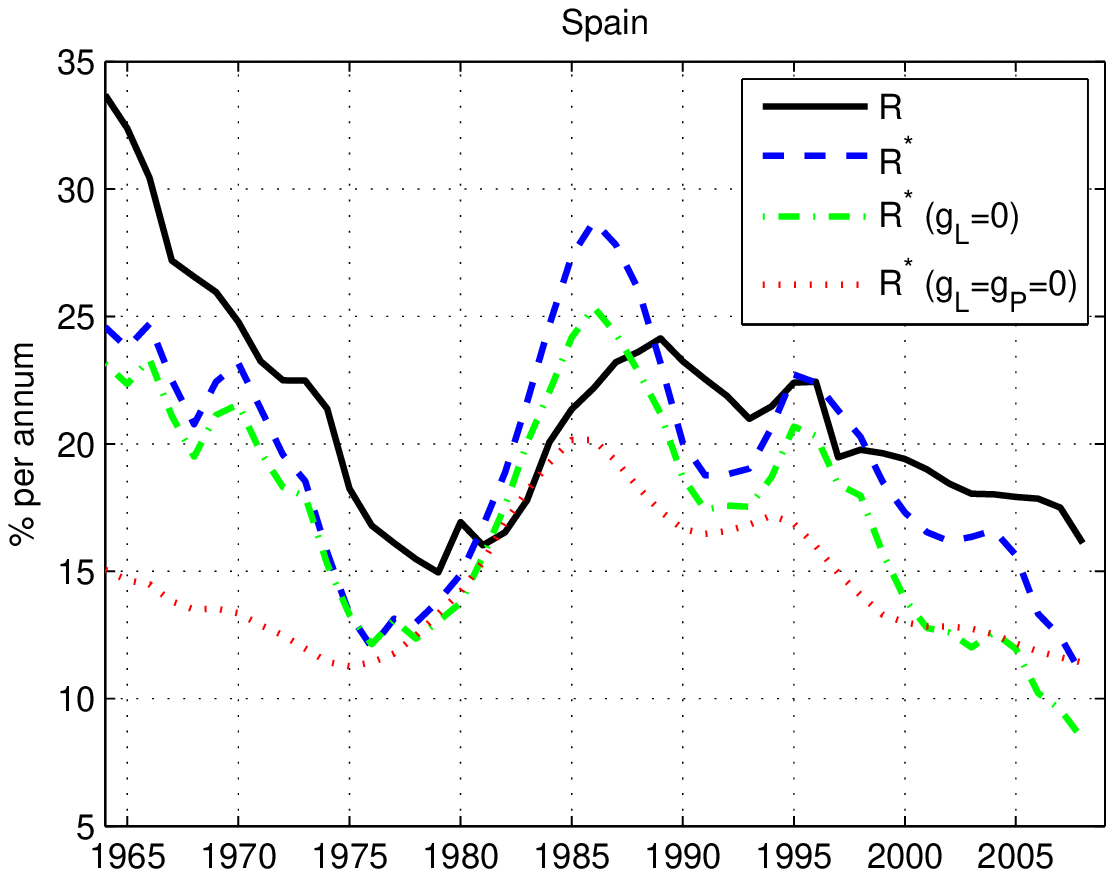}
        }%
        \subfigure{%
            \includegraphics[width=1.0\columnwidth]{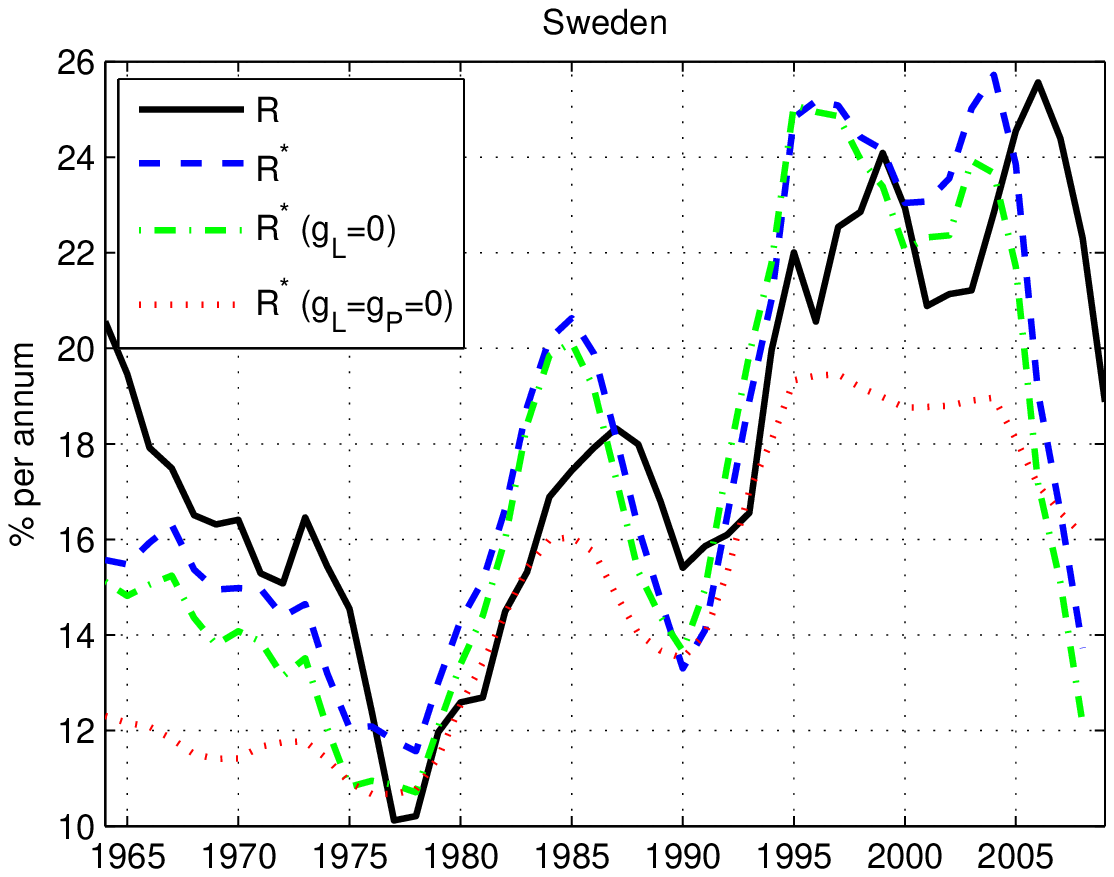}
        }%
    \end{center}
    \caption{Average profit rate $R$ and the equilibrium profit rate $R^*$, in four major Western European economies. Note that the contribution from the growth of labour $g_L$ is low or marginal. Source: \cite{Marquetti2012} and own calculations.}
   \label{fig:profitrate_EU}
\end{figure*}

We know that historically the process of industrialisation has combined
rapid accumulations of capital values with an exponential growth in
the working population. But we also know that societies undergo a
demographic transition once they have developed further -- with a shift
to zero or negative rates of natural population growth. As we have argued above, the end to
population growth is bound eventually to make the productive expansion of capitalism unviable. Figure~\ref{fig:profitrate_EU} illustrates the trajectories of the average profit rates, and their determinants, in the advanced economies of Western Europe.

If we look at Western Europe and Japan, the natural rate of population
growth is very low. In the face of this constraint we see a set of displacement
processes:
\begin{enumerate}
\item Pressure of immigration from areas of high birth rate. On the one
hand this is what one would expect simply as a diffusion process,
but there is also political `pressure' by capitalists interests to
reduce barriers to movement of labour in order to allow continued
capital growth.
\item As it becomes harder to reinvest profits in new labour employing activities,
there is an increasing dislocation between the apparent accumulation
of capital in financial instruments and the reality, in which the surplus
product is actually being socially used by the state.
\end{enumerate}
There is a fundamental conflict between the attempt to maintain a
mechanism of symbolic appropriation of the surplus by private owners
and the European demographic conditions which mean that little real
accumulation is now possible in the private sector. The greater part
of the social surplus product now has really to be appropriated by
the state as representative of society. The continuation of a private
claim on this surplus becomes actually infeasible. The conflict between
an exponential growth of financial claims and a stagnant population
upon whom those claims rest arguably lies at the heart of the `Euro
Crisis'.

\subsection{Greek debt}

For example we know that the Greek trade deficit with Germany and
the German trade surplus with Greece must sum to zero. It is impossible
to reduce one without reducing the other. If austerity in Greece reduces
their deficit with Germany, there must be a reduction in German exports.
If the objective is to reduce the trade imbalance between Germany
and Southern Europe, it would be better to give Germans longer and
better paid holidays so that they could spend more on Mediteranean
holidays rather than impoverishing both sides.

It is also helpful to look at the problem of state debts this way.
If at the end of a period the state is to reduce its liabilities and
improve its net worth, this necessarily implies a reduction in the
assets held by the state's creditors. How to reduce state debt and
how to impoverish the holders of state debt are one and the same problem.
In this light, a choice of options become visible
\begin{itemize}
\item debts can simply be repudiated,
\item inflation can be engineered to reduce the real value of the debt,
\item firms and individuals holding state bonds can be relieved of them
by the taxman,
\end{itemize}
In contrast to these direct measures, austerity imposed on the non-bondholding
classes, is much less effective. Since liabilities between sectors
of an economy must sum to zero, taxing those too poor to save will
only reduce state debt to the extent that the poor are driven to take
out increased personal loans. The increase in the liabilities of the
poor then compensates for the fall in state debt. The poor, however,
are not very credit worthy as the sub-prime mortgage collapse showed.

\subsection{The Stability Pact}

The key measures to solve the Euro crisis according to the treaty
recently adopted by most of the EU are to establish a balanced budget
or a budget surplus for the state sector.
\begin{quote}
BEARING IN MIND that the need for governments to maintain sound and
sustainable public finances and to prevent a general government deficit
becoming excessive is of essential importance to safeguard the stability
of the euro area as a whole, and accordingly, requires the introduction
of specific rules, including a `balanced budget rule'
and an automatic mechanism to take corrective action; \cite[p.2]{TSCG2012}

The Contracting Parties shall apply the rules set out in this paragraph
in addition and without prejudice to their obligations under European
Union law: (a) the budgetary position of the general government of
a Contracting Party shall be balanced or in surplus; \cite[article 3]{TSCG2012}
\end{quote}
The drafters of this treaty imagine that it is possible, by an act
of international law to impose a unilateral constraint on one item
in a mutually dependent complex of relations. The surplus or deficit
position of the government sector in the Eurozone depends upon the
net positions of all other sectors.

In Table \ref{tab:Sectoral-balances-for} we reproduce in summary
form the latest data on the surplus deficit positions of all financial
sectors in the Eurozone. All sectors other than the state sector run
a surplus, i.e., they are building up their financial assets. The change
proposed by the treaty is drastic. It proposes in effect to wipe out
all net financial transactions between the sectors by changing the
net borrowing position of the general government sector from a quarterly
deficit of 122 billion Euros to a deficit of 0 or even a net surplus.
\begin{table*}
\begin{center}
 \caption{Sectoral balances in billion Euros for the Eurozone, 2012 Q1, extracted from the online
database of the ECB on 4 Aug 2012. http://www.ecb.int/stats/money/aggregates/sect/html/index.en.html}
\label{tab:Sectoral-balances-for}
\begin{tabular}{rrrrrrrr}
 & Households & Nonfinancial & Financial & Govt. & Rest of &  & \\
 &  & companies & companies &  & world &  & \tabularnewline \hline
Net surplus  & 48 & 22 & 39 & --122 & 13 &  & \tabularnewline
 &  &  &  &  &  &  & \tabularnewline
\end{tabular}
\end{center}
\end{table*}

But the implication of this would be that the net savings of the household
sector, the company sector and the rest of the world would have to
be reduced to 0, effectively eliminating out all net financial transactions
as they exist today.

In the abstract this is possible. If by some means the savings of
the household sector could be completely eliminated, if the whole
company sector could be made to run at a break even position with
no net financial surplus, and if the Eurozone's trade deficit with
the rest of the world could be wiped out, then the government sector
could run a balanced budget.

At the level of accounting it is possible, but is such a measure
compatible with the continuation of a functioning capitalist economy?

Suppose the governments attempt to achieve this by austerity measures -- essentially cutting public expenditure. How does this affect each
of the other sectors?

\subsubsection{Household sector}

The `household sector', is an amalgam of different social classes.
It includes households from the propertied classes who are wealthy
enough to have a substantial financial surplus, but it includes far
more households who have little or no savings and are more likely
to be net debtors. The neo-liberal government austerity measures in
the EU today primarily target those on low incomes who have no financial
surplus. As such they only have a slight impact on the fiancial surplus
of the household sector.

In principle, austerity measures that would reduce the financial surplus
of the household sector are possible: for instance steep increases in income tax
on higher incomes or a progressive tax on large houses and landed
property. The Stafford Cripps austerity policies in the late 1940s
were effective in this way. Since such policies, e.g. top rates of income
tax above 90\%, are not being followed, the prospect of eliminating
the financial surplus of the personal sector is negligible.

\subsubsection{Nonfinancial companies}

In principle it would be possible, for a while at least, for
nonfinancial companies as a whole to run without a financial surplus,
or even with a financial deficit, if they were carrying out a large
programme of credit financed capital investment. But this scenario is
not very likely. Investment by nonfinancial companies tends to be
self-financing taken across the sector as a whole. Investment by one
company may require external funding, but its purchases boost the
profits of suppliers, which means that the sector as a whole tends to
self-finance.

Firms will only voluntarily seek external finance to expand if they
anticipate a high rate of profit on investment, which implies among
other things a rapidly expanding market for their products. It is
hard to see how this expanding market can be anticipated during a
period in which austerity measures are curtailing consumer demand.
Beyond that,   the rate of profit in Europe
has only been held up since the 1980s by a reduction in the accumulation
rate. An increased investment rate would thus be self-curtailing.

Involuntary deficits by industrial and commercial firms are of course
possible in the short run when faced with a slump in demand. But their
response to this is likely to be to quickly cut costs by shedding
labour, so even involuntary deficits induced by austerity would be
short lived. Attempting to force industrial and commercial companies
as a whole to run at break even point, which the treaty implies, would
mean, given the spread of rates of return within the sector, putting
a significant fraction of them on the path to bankruptcy. This again
is not sustainable.

\subsubsection{Financial companies}

Table \ref{tab:Sectoral-balances-for} shows that on an annual basis
financial companies in the Euro-zone are running a surplus of some
160 billion Euros and that they are thus responsible for $\frac{1}{3}$
of the total deficit of the general government sector.

There is almost nothing that the individual nation state governments
can do to eliminate this surplus with the framework of the Euro-zone.
Indeed the whole thrust of economic policy in the capitalist world
since the banking crisis broke out in 2008 has been to protect the
interests of financial companies. They could of course levy heavy
taxes on financial firms, but this is greatly complicated by the location
of the firms. The countries in the Euro-zone whose governments are
in the worst financial position are not necessarily the ones whose
financial firms are running the biggest surpluses.

A general reduction of interest rates would cut the surplus of the
financial sector, but that, under Euro system, is outwith the power
of the nation states. Only the ECB could systematically force down
interest rates by buying up national and local governent bonds. This
would reduce the cost of re-financing and over time would, to an
extent, reduce the financial surplus of the private banking sector.

The whole structure of the treaties governing the ECB has been
designed by the financial sector to prevent the ECB from acting in this way.

\subsubsection{Rest of the world}

The surplus of the rest of the world with the Euro-zone might be reduced
and even eliminated by neo-liberal austerity measures. A sufficiently
strong recession in the Euro-zone might cut imports sufficiently to
eliminate the trade deficit of the zone with the rest of the world.
But this of course would only succeed to the extent that the rest
of the world itself did not slip deeper into recession.

One effective tool that national governments have traditionally been
able to exercise is now out of reach for the Eurozone. They can no
longer devalue to bring their trade back into balance. The ECB again
could force a devaluation were it to act as the Bank of China used
to, and systematically buy up large quantities of dollar securities.
But the national governments can not instruct it to do so.

\subsection{Conclusion on the Stability Pact}

The structure set up under monetary union effectively makes it impossible
for national governments to meet the obligations that they have undertaken
in the pact. Any serious attempt to impose balanced budgets by austerity
measures will be ineffective in its professed aim, and would as a
side effect engender a downward spiral of bankruptcies, rising unemployment
and deepening economic ruin.

\bibliographystyle{plain}
\bibliography{ref_lawsofmotion}

\end{document}